\documentclass[10pt]{article}
\usepackage{colordvi}
\usepackage{epsfig}
\usepackage{axodraw}
\usepackage{epsfig}
\usepackage{graphicx}
\usepackage{rotate}
\usepackage{latexsym}
\usepackage{amssymb}
\usepackage[reqno,tbtags]{amsmath}
\usepackage{multirow}

\allowdisplaybreaks
\newcommand\pubnumber{ 
                       TTP      10-04  }
\newcommand\pubdate{January 19, 2010}

\def\csumb{Dipartimento di Fisica Teorica, Universit\`a di Torino, Italy\\
           INFN, Sezione di Torino, Italy}
\def\csumc{Physics Department, Brookhaven National Laboratory,\\
           Upton, NY 11973, USA}
\def\csumd{Institut f\"ur Theoretische Teilchenphysik, Universit\"at Karlsruhe,\\
           76128 Karlsruhe, Germany}

\def\Title#1{\begin{center} {\Large\bf #1 } \end{center}}

\def\Author#1{\begin{center}{ \sc #1} \end{center}}
\def\Address#1{\begin{center}{ \it #1} \end{center}}

\newcommand\pubblock{\rightline{\begin{tabular}{l} \\
         \pubnumber\\ \\ \pubdate\\  \end{tabular}}}
\newenvironment{Abstract}{\begin{quotation}  }{\end{quotation}}

\def\Acknowledgments{\bigskip  \bigskip \begin{center}
          \large\bf Acknowledgments\end{center}}
\def\email#1{\footnote{#1}}
\makeatletter
\def\section{\@startsection{section}{0}{\z@}{5.5ex plus .5ex minus
 1.5ex}{2.3ex plus .2ex}{\large\bf}}
\def\subsection{\@startsection{subsection}{1}{\z@}{3.5ex plus .5ex minus
 1.5ex}{1.3ex plus .2ex}{\normalsize\bf}}
\def\subsubsection{\@startsection{subsubsection}{2}{\z@}{-3.5ex plus
-1ex minus  -.2ex}{2.3ex plus .2ex}{\normalsize\sl}}

\renewcommand{\@makecaption}[2]{%
   \vskip 10pt
   \setbox\@tempboxa\hbox{\small #1: #2}
   \ifdim \wd\@tempboxa >\hsize     
       \small #1: #2\par          
     \else                        
       \hbox to\hsize{\hfil\box\@tempboxa\hfil}
   \fi}
 \def\citenum#1{{\def\@cite##1##2{##1}\cite{#1}}}
\def\citea#1{\@cite{#1}{}}
\newcount\@tempcntc
\def\@citex[#1]#2{\if@filesw\immediate\write\@auxout{\string\citation{#2}}\fi
  \@tempcnta\z@\@tempcntb\m@ne\def\@citea{}\@cite{\@for\@citeb:=#2\do
    {\@ifundefined
       {b@\@citeb}{\@citeo\@tempcntb\m@ne\@citea\def\@citea{,}{\bf }\@warning
       {Citation `\@citeb' on page \thepage \space undefined}}%
    {\setbox\z@\hbox{\global\@tempcntc0\csname b@\@citeb\endcsname\relax}%
     \ifnum\@tempcntc=\z@ \@citeo\@tempcntb\m@ne
       \@citea\def\@citea{,}\hbox{\csname b@\@citeb\endcsname}%
     \else
      \advance\@tempcntb\@ne
      \ifnum\@tempcntb=\@tempcntc
      \else\advance\@tempcntb\m@ne\@citeo
      \@tempcnta\@tempcntc\@tempcntb\@tempcntc\fi\fi}}\@citeo}{#1}}
\def\@citeo{\ifnum\@tempcnta>\@tempcntb\else\@citea\def\@citea{,}%
  \ifnum\@tempcnta=\@tempcntb\the\@tempcnta\else
  {\advance\@tempcnta\@ne\ifnum\@tempcnta=\@tempcntb \else\def\@citea{--}\fi
    \advance\@tempcnta\m@ne\the\@tempcnta\@citea\the\@tempcntb}\fi\fi}
\makeatother
\newcommand{\nl}{\nonumber\\}
\newcommand{\nn}{\nonumber}

\newcommand{\lpar}{\left(}                            
\newcommand{\rpar}{\right)}

\newcommand{\bq}{\begin{equation}}                    
\newcommand{\eq}{\end{equation}}
\newcommand{\bqa}{\arraycolsep 0.14em\begin{eqnarray}}
\newcommand{\eqa}{\end{eqnarray}}
\newcommand{\ba}[1]{\begin{array}{#1}}
\newcommand{\ea}{\end{array}}
\newcommand{\ben}{\begin{enumerate}}
\newcommand{\een}{\end{enumerate}}
\newcommand{\bei}{\begin{itemize}}
\newcommand{\eei}{\end{itemize}}
\newcommand{\eqn}[1]{Eq.(\ref{#1})}

\newcommand{\tabn}[1]{Tab.~\ref{#1}}

\newcommand{\fig}[1]{Fig.~\ref{#1}}

\newcommand{\sect}[1]{Section~\ref{#1}}
\newcommand{\sects}[2]{Section~\ref{#1} and \ref{#2}}

\newcommand{\GeV}{\mathrm{GeV}}

\def\Re{\mathop{\operator@font Re}\nolimits}
\def\Im{\mathop{\operator@font Im}\nolimits}
\newcommand{\ord}[1]{{\cal O}\lpar#1\rpar}

\newcommand{\barb}{\overline b}
\newcommand{\bart}{\overline t}

\newcommand{\baru}{\overline u}

\newcommand{\mw}{M_{_W}}

\newcommand{\mz}{M_{_Z}}

\newcommand{\mh}{M_{_H}}

\newcommand{\mt}{m_t}

\newcommand{\mws}{M^2_{_W}}

\newcommand{\mzs}{M^2_{_Z}}

\newcommand{\mhs}{M^2_{_H}}

\newcommand{\gf}{G_{\ssF}}

\newcommand{\stw}{s_{\theta}}             

\newcommand{\spro}[2]{{#1}\cdot{#2}}
\newcommand{\li}[2]{\mathrm{Li}_{#1}\lpar\displaystyle{#2}\rpar} 

\newcommand{\intfx}[1]{\int_{\scriptstyle 0}^{\scriptstyle 1}\,d#1}
\newcommand{\intfxy}[2]{\int_{\scriptstyle 0}^{\scriptstyle 1}\,d#1\,
                        \int_{\scriptstyle 0}^{\scriptstyle #1}\,d#2}

\newcommand{\sign}[1]{{\rm{sign}}\lpar{#1}\rpar}

\newcommand{\ep}{\epsilon}

\newcommand{\Reb}{{\rm{Re}}}
\newcommand{\Imb}{{\rm{Im}}}
\newcommand{\upar}[1]{u}

\newcommand{\mhss}{M^2_{\ssH}}

\newcommand{\atan}[1]{{\rm{arctan}}\lpar{#1}\rpar}

\newcommand{\ssB}{{\scriptscriptstyle{B}}}
\newcommand{\ssC}{{\scriptscriptstyle{C}}}

\newcommand{\ssE}{{\scriptscriptstyle{E}}}
\newcommand{\ssF}{{\scriptscriptstyle{F}}}

\newcommand{\ssH}{{\scriptscriptstyle{H}}}
\newcommand{\ssI}{{\scriptscriptstyle{I}}}

\newcommand{\ssL}{{\scriptscriptstyle{L}}}

\newcommand{\ssP}{{\scriptscriptstyle{P}}}

\newcommand{\ssR}{{\scriptscriptstyle{R}}}
\newcommand{\ssS}{{\scriptscriptstyle{S}}}

\newcommand{\ssV}{{\scriptscriptstyle{V}}}
\newcommand{\ssW}{{\scriptscriptstyle{W}}}

\newcommand{\ssZ}{{\scriptscriptstyle{Z}}}

\newcommand{\QED}{\rm{\scriptscriptstyle{QED}}}

\newcommand{\EW}{\rm{\scriptscriptstyle{EW}}}

\newcommand{\NLO}{\rm{\scriptscriptstyle{NLO}}}
\newcommand{\LO}{\rm{\scriptscriptstyle{LO}}}
\newcommand{\CMRP}{\rm{\scriptscriptstyle{CMRP}}}
\newcommand{\CMCP}{\rm{\scriptscriptstyle{CMCP}}}

\newcommand{\bqas}{\begin{eqnarray*}}
\newcommand{\eqas}{\end{eqnarray*}}

\def\app#1#2 {{\it Acta. Phys. Pol.} {\bf#1},#2}
\def\cpc#1#2 {{\it Computer Phys. Comm.} {\bf#1},#2}
\def\np#1#2 {{\it Nucl. Phys.} {\bf#1},#2}
\def\pl#1#2 {{\it Phys. Lett.} {\bf#1},#2}
\def\prep#1#2 {{\it Phys. Rep.} {\bf#1},#2}
\def\prev#1#2 {{\it Phys. Rev.} {\bf#1},#2}
\def\prl#1#2 {{\it Phys. Rev. Lett.} {\bf#1},#2}
\def\zp#1#2 {{\it Zeit. Phys.} {\bf#1},#2}
\def\sptp#1#2 {{\it Suppl. Prog. Theor. Phys.} {\bf#1},#2}
\def\mpl#1#2 {{\it Modern Phys. Lett.} {\bf#1},#2}
\def\jetp#1#2 {{\it Sov. Phys. JETP} {\bf#1},#2}
\def\fpj#1#2 {{\it Fortschr. Phys.} {\bf#1},#2}
\def\afp#1#2 {{\it Acta.Phys. Polon.} {\bf#1},#2}
\def\err#1#2 {{\it Erratum} {\bf#1},#2}
\def\ijmp#1#2 {{\it Int. J. Mod. Phys} {\bf#1},#2}
\def\nc#1#2 {{\it Nuovo Cimento} {\bf#1},#2}
\def\ap#1#2 {{\it Ann. Phys.} {\bf#1},#2}
\def\cmp#1#2 {{\it Comm. Math. Phys.} {\bf#1},#2}
\def\el#1#2 {{\it Europhys. Lett.} {\bf#1},#2}
\def\hpa#1#2 {{\it Helv. Phys. Acta} {\bf#1},#2}
\def\yf#1#2 {{\it Yad. Fiz.} {\bf#1},#2}
\def\nim#1#2 {{\it Nucl. Instrum. Meth.} {\bf#1},#2}
\def\spz#1#2 {{\it Sov. Pisma Zhetf} {\bf#1},#2}
\def\jetpl#1#2 {{\it JETP Lett.} {\bf#1},#2}
\def\sjnp#1#2 {{\it Sov. J. Nucl. Phys.} {\bf#1},#2}
\def\ptp#1#2 {{\it Progr. Theor. Phys. (Kyoto)} {\bf#1},#2}
\def\rmp#1#2  {{\it Rev. Mod. Phys.} {\bf#1},#2}
\def\zhetf#1#2 {{\it ZhETF} {\bf#1},#2}
\def\prs#1#2 {{\it Proc. Roy. Soc.} {\bf#1},#2}
\def\phys#1#2 {{\it Physica} {\bf#1},#2}

\def\bfi{\begin{figure}}
\def\efi{\end{figure}}

\newcommand{\intsx}[1]{\int_{\scriptstyle 0}^{\scriptstyle 1}\!\!\!d#1}

\newcommand{\bmid}{\Bigr|}

\newcommand{\cph}{s_{\ssH}}

\newcommand{\rph}{\mu^2_{\ssH}}
\newcommand{\srph}{\mu_{\ssH}}
\newcommand{\lgh}{\gamma_{\ssH}}
\newcommand{\brph}{{\overline\mu}^2_{\ssH}}
\newcommand{\sbrph}{{\overline\mu}_{\ssH}}

\newcommand{\ssHH}{{\scriptscriptstyle{H}\scriptscriptstyle{H}}}

\newcommand{\oD}{{\overline \Delta}}

\newcommand{\ST}{\rm{\scriptscriptstyle{ST}}}
\newcommand{\UST}{\rm{\scriptscriptstyle{UST}}}

\begin{document}
\begin{titlepage}
\pubblock
%
\vfill
\def\thefootnote{\fnsymbol{footnote}}
\Title{Higgs Pseudo-Observables,\\[0.3cm]
Second Riemann Sheet and All That
  \footnote[9]{This work is supported by the European Community's Marie Curie Research 
               Training Network {\it Tools and Precision Calculations for 
               Physics Discoveries at Colliders} under contract 
               MRTN-CT-2006-035505, by the U.S. Department of Energy under 
               contract No. DE-AC02-98CH10886 and by the Deutsche 
               Forschungsgemeinschaft through Sonderforschungsbereich/Transregio 9 
               {\it Computergest\"utzte Theoretische Teilchenphysik}.}}
\vfill
\Author{Giampiero Passarino 
\email{giampiero@to.infn.it}}               
\Address{\csumb}
\Author{Christian Sturm     
\email{sturm@bnl.gov}}            
\Address{\csumc}
\Author{Sandro Uccirati     
\email{uccirati@particle.uni-karlsruhe.de}} 
\Address{\csumd}
\vfill
\vfill
\begin{Abstract}
\noindent 
The relation between physical observables measured at LHC and Tevatron and 
standard model Higgs pseudo-observables (production cross section and partial 
decay widths) is revised by extensively using the notion of the Higgs complex pole 
on the second Riemann sheet of the $S\,$-matrix. The extension of their definition 
to higher orders is considered, confronting the problems that arise when QED(QCD) 
corrections are included in computing realistic observables. Numerical results 
are presented for pseudo-observables related to the standard model Higgs boson 
decay and production. The relevance of the result for exclusion plots of the 
standard model Higgs boson for high masses (up to $600\,$GeV) is discussed. 
Furthermore, a recipe for the analytical continuation of Feynman loop integrals 
from real to complex internal masses and complex Mandelstam invariants is 
thoroughly discussed.
\end{Abstract}
\vfill
\begin{center}
\end{center}
\end{titlepage}
\def\thefootnote{\arabic{footnote}}
\setcounter{footnote}{0}
\small
\thispagestyle{empty}
\tableofcontents
\normalsize
\clearpage
\setcounter{page}{1}
\section{Introduction \label{intro}}
The search for a mechanism explaining electroweak symmetry breaking has been
a major goal for many years, in particular the search for a standard model (SM)
Higgs boson, see for instance Ref.~\cite{Phenomena:2009pt} and 
Ref.~\cite{Fernandez:2009ac}. As a result of this an intense effort in
the theoretical community has been made to produce the most accurate NLO
and NNLO predictions, see 
Refs.~\cite{Anastasiou:2008tj,Grazzini:2008zz,Actis:2008uh,Harlander:2009my}. 
There is, however, a point that has been ignored in all these calculations: the 
Higgs boson is an unstable particle and should be removed from the in/out bases in the
Hilbert space, without destroying the unitarity of the theory. Therefore,
concepts as the {\em production} of an unstable particle or its {\em partial decay 
widths} do not have a precise meaning and should be replaced by a
conventionalized definition which respects first principles of quantum field theory
(QFT). 

The quest for a proper treatment of a QFT of unstable particles dates back to
the sixties and to the work of Veltman~\cite{Veltman:1963th} (for earlier
attempts see Ref.~\cite{peierls}); more recently the question has been readdressed 
by Sirlin and collaborators~\cite{Grassi:2000dz}.
Alternative approaches, within the framework of an effective theory can be found 
in Ref.~\cite{Beneke:2003xh}.

In this paper we discuss the relation between physical observables and Higgs 
pseudo-observables by considering the extension of their definition to higher 
orders in perturbation theory, confronting the problems that arise when perturbative
corrections in quantum electrodynamics (QED) and quantum chromodynamics
(QCD) are included. Numerical results are also presented. 
Our work can be seen as an extension of complex-mass schemes to include complex
external momenta (for previous work see also Ref.~\cite{Valent:1974bd}), 
addressing systematically the question of the analytical continuation of Feynman 
loop integrals.

This paper is organized as follows. In \sect{theP} we summarize the conceptual 
setup. In \sect{CP} we present general arguments on complex poles.
In \sects{PDW}{UNI} we discuss pseudo-observables, on-shell observables and 
unitarity. The analytical continuation of Feynman 
loop integrals into the second Riemann sheet of the $S\,$-matrix is
examined in \sect{allcmplx}.
In \sect{theQs} we present the inclusion of QED and QCD corrections and
renormalization schemes are highlighted in \sect{schemes}.
Numerical results are given in \sect{Nres} and in \sect{Conclu} we close
with our conclusions.
\section{Formulation of the problem \label{theP}}
There are two old questions in relating measurements to theoretical predictions:
\begin{itemize}

\item[--] Experimenters (should) extract so-called {\em realistic observables} from
raw data, e.g. $\sigma (p p \to \gamma \gamma + X)$ and need to present
results in a form that can be useful for comparing them with theoretical
predictions, i.e. the results should be transformed into pseudo-observables;
during the deconvolution procedure one should also account for the interference 
background -- signal;

\item[--] Theorists (should) compute pseudo-observables using the best available
technology and satisfying a list of demands from the self-consistency of
the underlying theory~\cite{Bardin:1999gt}.

\end{itemize}

Almost from the start it is clear that a common language must be established 
in order to avoid misunderstandings and confusion. A typical example can be found 
in Higgs physics where, frequently, one talks about {\em Higgs production
cross section} or {\em Higgs partial decay widths}. 
After the discovery phase, in absence of which the future of high energy
physics cannot be ascertained, one will need to probe the properties
of the discovered resonance, like spin and couplings. In this case
different sources will start talking about the same thing but with 
different languages%
. We will indicate a reasonable language within the context of a
perturbative expansion of a gauge-invariant QFT in this paper.  

The Higgs boson, as well as the $W$ or $Z$ bosons, are unstable 
particles; as such they should be removed from in/out bases in the
Hilbert space, without changing the unitarity of the theory. 
As mentioned before, concepts as the production of an unstable particle 
or its partial decay widths, not having a precise meaning, are only an 
approximation of a more complete description.
The inconsistencies associated with the on-shell LSZ formulation of an
unstable external particles become particularly severe starting from two-loops,
as described in Ref.~\cite{Actis:2008uh}.

Suppose that we want to combine a Higgs production mechanism, say 
gluon-gluon fusion, with the subsequent decay $H \to \gamma \gamma$. The 
process to be considered is, therefore, $pp \to \gamma \gamma + X$
and it is made of a part that defines the signal, e.g.
\bq
pp \to g g ( \to H \to \gamma \gamma ) + X,
\eq
and by a non-resonant background. The question is: how to extract
from the data, without ambiguities, a pseudo-observable to be termed
{\em Higgs partial decay width into two photons} which, at the same time,
does not violate first principles? Once again, one should be aware that there 
is no Higgs boson in the in-state, therefore the matrix element 
$<\gamma \gamma\; {\rm out} | H\; {\rm in}>$ is not definable in QFT and 
this ill-defined quantity should be replaced by a pseudo-observable which
closely resembles the intuitive concept of a decay width, can be
unambiguously extracted from the data and respects all fundamental 
properties of the theory; in this way we replace a {\em non existing} 
observable with a conventional definition.
A proposal in this direction can be found in Ref.\cite{Grassi:2000dz}; 
here we revise the proposal, improving it by considering the extension 
to higher orders in perturbation theory, confronting the problems that 
arise when QED(QCD) corrections have to be included and present numerical 
results for Higgs physics.

At the parton level the $S\,$-matrix for the process $i \to f$ can be written 
as
\bq
S_{fi} = V_i(s)\,\Delta_{\ssH}(s)\,V_f(s) + B_{if}(s),
\label{Smat}
\eq
where $V_i$ is the production vertex $i \to H$ (e.g. $gg \to H$), $V_f$ 
is the decay vertex $H \to f$ (e.g. $H \to \gamma \gamma$), $\Delta_{\ssH}$ 
is the Dyson re-summed Higgs propagator and $B_{if}$ is the non-resonant 
background (e.g. $gg \to \gamma \gamma$ boxes). In the next section we 
will introduce the notion of complex pole. 
A vertex is defined by the following decomposition~\cite{Grassi:2001bz},
\bq
V_f(s) = \sum_a\,V^a_f\lpar s\,,\,\{S\}\rpar\,F^a_f\lpar \{p_f\}\rpar
\eq
where $s = - P_{\ssH}^2$ (with $P_{\ssH} = \sum_f p_f$), $s\,\oplus\,\{S\}$ 
is the set of Mandelstam invariants that characterize the process 
$H \to f$, $V^a_f$ are scalar form factors and the $F^a_f$ contain spinors, 
polarization vectors, etc.  
\section{The complex pole \label{CP}}
In this section we introduce the notion of the complex pole~\cite{Stuart:1991xk}
following closely the analysis of Ref.~\cite{Actis:2006rc}.
Let $\Delta_i$ be the lowest order propagator for particle $i$ and $\oD_i$ the 
corresponding dressed propagator, i.e.
\bq
\oD_i = -\,\frac{\Delta_i}{1 + \Delta_i\,\Sigma_{ii}},
\eq
Let us analyze in more details the definition of the dressed propagator: 
to begin with, consider a skeleton expansion of the self-energy 
$S= 16\,\pi^4\,i\,\Sigma$ with propagators that are resummed up to $\ord{n}$ and define
\bq
\Delta^{(n)}_i(s) = -\,\Delta^{(0)}_i(s)\,\Bigl[
1 + \Delta^{(0)}_i(s)\,\Sigma^{(n)}_{ii}\lpar s\,,\,
\Delta^{(n-1)}_i(s)\rpar\Bigr]^{-1},
\eq
where, omitting an overall factor $-i/(2\,\pi)^4$, the Born propagator 
(tensor structures are easily included) is
\bq
\Delta^{(0)}_i(s) = \frac{1}{s - m^2_i}.
\eq
If it exists, we define a dressed propagator as the formal 
limit~\cite{Veltman:1963th}
\bq
\oD_i(s) = \lim_{n \to \infty}\,\Delta^{(n)}_i(s),
\qquad
\oD_i(s) =
-\,\Delta^{(0)}_i(s)\,\Bigl[
1 + \Delta^{(0)}_i(s)\,\Sigma_{ii}\lpar s\,,\,\oD_i(s)\rpar\Bigr]^{-1},
\label{dr}
\eq 
which coincides with the Schwinger-Dyson solution for the propagator.

The Higgs boson complex pole ($\cph$) is the solution of the equation
\bq
\cph - \mhss + \Sigma_{\ssHH}(\cph) = 0,
\label{eqCP}
\eq
where $\mhss$ is the renormalized Higgs boson mass; here we assume that all
counter-terms have been introduced to make the off-shell Green's function
ultraviolet finite, respecting locality of the counter-terms. 
We now examine more carefully the self-energy to all orders in perturbation 
theory since, often, there is some confusion with statements that are 
formulated to all orders and applied to a truncated perturbative expansion. 
Now consider the, all orders, self-energy,
\bq
\Sigma_{\ssHH}(s,\mhss,\xi) = \sum_{n=1}^{\infty}\,
\Sigma^{(n)}_{\ssHH}(s,\mhss,\xi)\,g^{2n},
\eq
where $\xi$ is the gauge parameter (extension to more than one gauge 
parameters is straightforward) and $g$ is the renormalized coupling constant.
From arguments based on Nielsen identities, see Ref~\cite{Grassi:2001bz}, we 
know that
\bq
\frac{\partial}{\partial \xi}\,\cph =  0,
\qquad
\frac{\partial}{\partial \xi}\,\Sigma_{\ssHH}(\cph,\mhss,\xi) =  0,
\label{NI}
\eq
i.e. the location of the complex pole is $\xi$ independent; as a 
consequence the self-energy is $\xi$ independent too, since the two differ 
by a renormalized quantity, obviously $\xi$ independent.
We consider first the one-loop approximation for the self-energy: 
from its explicit expression we are able to derive the following relation:
\bq
\Sigma^{(1)}_{\ssHH}(s,\mhss,\xi) =
\Sigma^{(1)}_{\ssHH\,;\,\ssI}(s,\mhss) +
( s - \mhss)\,\Phi_{\ssH}(s,\mhss,\xi).
\label{olSE}
\eq
where, in a general $R_{\xi}$ gauge, we obtain
\bq
\mws\Phi_{\ssH} = 
-\frac{1}{8}\bigg\{ 
    (s \!+\! \mhss)\Bigl[ 
    B_d(s,\mzs,\mzs; \xi_{\ssZ}) + 2\, B_d(s,\mw,\mw; \xi_{\ssW}) 
    \Bigr]
  + 2\,A_d(\mz,\xi_{\ssZ}) + 4\,A_d(\mw,\xi_{\ssW})
\bigg\},
\nn
\eq
\bq
B_d\lpar s,m,m,\xi\rpar = 
B_0\lpar s,\xi\,m,\xi\,m\rpar - B_0\lpar s,m,m\rpar,
\quad
A_d\lpar m,\xi \rpar = A_0(\xi\,m) - A_0(m). 
\eq
The symbols $A_0, B_0,\,$ etc. are the usual scalar, one-loop functions.

It needs to be stressed that the splitting between gauge dependent and gauge 
independent quantities is only defined modulo a $\xi\,$-independent constant. 
Our definition of the invariant part is that it coincides with the 
expression in the 't Hooft-Feynman gauge (i.e. $\xi = 1$).
Furthermore, finite renormalization (i.e. replacing renormalized 
parameters with a set of {\em experimental} data points after having 
removed ultraviolet poles by means of local counter-terms) amounts to 
replace 
\bq
\mhss = \cph + \Sigma_{\ssHH}(\cph,\mhss,\xi),
\label{CPmren}
\eq
showing that
\bq
\frac{\partial}{\partial \xi}\,
\Sigma^{(1)}_{\ssHH}(\cph,\cph,\xi) = 0,
\eq
so that, at one-loop, the Higgs complex pole is gauge parameter independent
if the self-energy is computed at $\mhss = \cph$, the basis of the
so-called complex-mass scheme (see Ref.~\cite{Denner:2005fg} and also
Ref.~\cite{Actis:2006rc}). 
From \eqn{NI} and from the one-loop result in \eqn{olSE}, we derive the 
following
\bqa
\Sigma^{(n)}_{\ssHH}(\cph,\mhss,\xi) &=& 
\Sigma^{(n)}_{\ssHH\,;\,\ssI}(\cph,\mhss) + 
\Sigma^{(n)}_{\ssHH\,;\,\xi}(\cph,\mhss,\xi),
\nl  
\Sigma^{(n)}_{\ssHH\,;\,\xi}(\cph,\mhss,\xi) &=& 
\Sigma^{(n-1)}_{\ssHH\,;\,\ssI}(\cph,\mhss)\,
\Phi_{\ssH}(\cph,\mhss,\xi).
\label{niII}
\eqa
Using \eqn{niII} we can rewrite \eqn{eqCP} at the two-loop level as
\bq
\mhss = 
  \cph 
+ g^2\,\Sigma^{(1)}_{\ssHH\,;\,\ssI}(\cph,\cph) 
+ g^4\Bigl[ 
  \Sigma^{(2)}_{\ssHH\,;\,\ssI}(\cph,\cph) +
  \Sigma^{(1)}_{\ssHH\,;\,\ssI}(\cph,\cph)\,
  \frac{\partial}{\partial \mhss}\,
  \Sigma^{(1)}_{\ssHH\,;\,\ssI}(\cph,\mhss)\bmid_{\mhss=\cph}
  \Bigr].
\label{eqCPtl}
\eq
We find as worthy note that \eqn{eqCPtl} can be easily generalized to all 
orders in perturbation theory showing that, order-by-order, the gauge 
dependent part of the self-energy drops out in the equation for the complex 
pole of the particle. The complex pole, sitting on the second Riemann sheet of 
the $S\,$-matrix, is usually parametrized as
\bq
\cph = \rph - i\,\srph\,\lgh.
\eq
It is worth noting that a consistent treatment of external ($s$) and 
internal ($\mhss$) masses allows the extension of the complex mass 
scheme beyond one-loop, without the need of expanding the self-energy 
around $\cph = \rph$, as frequently done in the literature.
In partial contrast to the traditional complex mass scheme, 
Ref.~\cite{Denner:2005fg}, in our approach (described in 
Ref.~\cite{Actis:2006rc}) it is the finite renormalization equation and 
not the Lagrangian that is modified.
Indeed, calling the scheme {\em complex mass scheme} is somehow misleading; 
to the requested order we replace everywhere the renormalized mass 
$M^2_{\ssB}$ with $s_{\ssB} + \Sigma_{\ssB\ssB}(s_{\ssB})$ which is real by 
construction;
if only one-loop is needed then $M^2_{\ssB} \to s_{\ssB}$ everywhere, therefore
justifying the name {\em complex mass}.

The quest for gauge invariance and the consequent introduction of a 
complex pole instead of an on-shell mass signal has a certain degree of 
ambiguity in defining the Higgs boson mass (as well as the mass of any 
unstable particle). 
The most convenient choice, for all practical purposes, is represented 
by the square root of the real part of $\cph$, although 
\bq
\brph = \srph\,\lpar \rph + \gamma^2_{\ssH}\rpar^{1/2}
\label{MHdef}
\eq
also has several advantages~\cite{Ghinculov:1996py} and will be used in
our numerical results. 

There is a final comment for this section: the complex pole for an unstable 
particle, parametrized according to \eqn{eqCP}, must correspond to a negative 
imaginary part; otherwise, even the Wick rotation cannot be safely performed. 
Consider the case of the Higgs boson, $ii$ channels that do not 
satisfy the negativity condition for the imaginary part below the $4\,m^2_i$ 
(real) threshold are excluded in the evaluation of $\cph$. 
As we already mentioned the contribution to the imaginary part of $\cph$ 
from a given channel below the corresponding real threshold ($WW$, $ZZ$ 
and $\bart t$) represents an approximation to the corresponding $4f$ and 
$6f$ cuts, i.e. $H \to WW, ZZ \to 4f$ etc, which is acceptable only when 
the corresponding $\lgh$ is positive, a condition which fails at one-loop 
for $\bart t$ intermediate states when the top quark mass is kept real; 
in this case $\bart t$ intermediate states never contribute, in our scheme, 
to $\lgh$ below threshold,  i.e. they are discarded. 
It is interesting to note that this problem completely disappears if we allow 
for a top quark complex pole (instead of real on-shell mass). 
Numerical examples will be discussed in \sect{Nres}; unfortunately the top 
quark total (on-shell) width is poorly known, therefore inducing large 
uncertainties on the corrections. 
In the numerical analysis we use $\Gamma_t \le 13.1\,$GeV, based on the 
experimental upper limit of Ref.~\cite{Aaltonen:2008ir}.
\section{Extracting a partial decay width \label{PDW}}
In this section we examine our options to define a pseudo-observable 
which is related, as closely as possible, to a realistic cross section 
and shares as many features as possible with the corresponding on-shell 
definition of a partial decay width. 
If we insist that $|H>$ is an asymptotic state in the Hilbert space 
then the observable to consider will be $<f\,{\rm out}\,|\,H\,{\rm in}>$, 
otherwise one should realize that for stable particles the proof of the LSZ 
reduction formulas depends on the existence of asymptotic states
\bq
|\,p\,{\rm in}\, > = \lim_{t \to -\,\infty}\int\,d^3 x\,H(x)\,i\,
{\partial_t}\!\!\!\!\!^{^{\leftrightarrow}} \,
e^{i\,\spro{p}{x}}\,|\,0\,>,
\eq
(in the weak operator sense). 
For unstable particles the energy is complex so that this limit either 
diverges or vanishes. 
Although a modification of the LSZ reduction formulas has been proposed long
ago for unstable particles, see Ref.~\cite{Weldon:1975gu}, we prefer an
alternative approach where one considers extracting information on the Higgs 
boson directly from
\bq
<\,f\;{\rm out}\,|\,H\,>\,<\,H\,|\,i\;{\rm in}\,> +
\sum_{n\,\not=\,H}\,
<\,f\;{\rm out}\,|\,n\,>\,<\,n\,|\,i\;{\rm in}\,>,
\eq
for some initial state $i$ and some final state $f$ and where 
$\{n\}\,\oplus\,H$ is a complete set of states (not as in the in/out bases). 
As we are about to see, the price to be paid is the necessity of moving into 
the complex plane. Define $\Pi_{\ssHH}(s)$ as
\bq
\Pi_{\ssHH}(s) = \frac{\Sigma_{\ssHH}(s) - \Sigma_{\ssHH}(\cph)}{s-\cph},
\eq
then the, Dyson re-summed, Higgs propagator becomes
\bq
\Delta_{\ssHH}(s)= (s - \cph)^{-1}\,\Bigl[ 1 + \Pi_{\ssHH}(s)\Bigr]^{-1},
\qquad
Z_{\ssH} = 1 + \Pi_{\ssHH}.
\label{tres}
\eq
Using \eqn{tres} we can write \eqn{Smat} as
\bq
S_{fi} = \Bigl[ Z^{-1/2}_{\ssH}(s)\,V_i(s)\Bigr]\,
            \frac{1}{s - \cph}\,
         \Bigl[ Z^{-1/2}_{\ssH}(s)\,V_f(s)\Bigr] + B_{if}(s).
\eq
From the $S\,$-matrix element for a physical process $i \to f$ we extract the 
relevant pseudo-observable,
\bq
S\lpar H_c \to f\rpar = Z^{-1/2}_{\ssH}(\cph)\,V_f(\cph),
\label{PO}
\eq
which is gauge parameter independent -- by construction -- and satisfies the
relation
\bq
S_{fi} = \frac{S\lpar i \to H_c \rpar\,S\lpar H_c \to f \rpar}{s - \cph} +
\hbox{non resonant terms}.
\eq
The partial decay width is further defined as
\bq
\srph\,\Gamma\lpar H_c \to f\rpar = \frac{(2\,\pi)^4}{2}\,\int\,
d\Phi_f\lpar P_{\ssH}\,,\,\{p_f\}\rpar\,
\sum_{\rm spins}\,\bmid S\lpar H_c \to f\rpar \bmid^2,
\label{GPO}
\eq
where the integration is over the phase space spanned by $| f >$, with the
constraint $P_{\ssH} = \sum\,p_f$. One should not confuse phase space and
the real value of $s= -P^2_{\ssH}$, where the realistic observable is 
measured, with the complex value for $s$, where gauge invariant 
loop corrections must be computed.
The choice of $P^2_{\ssH}$ (phase space) where to define the 
pseudo-observable is conventional, e.g. one can use the real part 
of $\cph$. Indeed, the r.h.s. of \eqn{PO} satisfies the property
\bq
\frac{\partial}{\partial \xi}\,Z^{-1/2}_{\ssH}(\cph)\,V_f(\cph) = 0
\eq
to all orders in perturbation theory. If we define
\bq
V_f\lpar s,\mhss\rpar = \sum_{n=0}^{\infty}\,g^{2 n + 1}\,\Bigl[ 
V^{(n)}_{f\,;\,\ssI}\lpar s,\mhss\rpar + 
V^{(n)}_{f\,;\,\xi}\lpar s,\mhss\rpar \Bigr],
\eq
we obtain, expanding in powers of the coupling constant $g$, that
\bq
V^{(1)}_{f;\,\xi}(\cph,\cph) =
\frac{1}{2}\,V^{(0)}\Phi_{\!\ssH}(\cph,\cph),
\qquad
V^{(2)}_{f;\,\xi}( \cph,\cph) =
- \frac{1}{2}\Phi_{\!\ssH}(\cph,\cph)\Bigl[
    V^{(1)}_{f;\,\xi}( \cph,\cph) 
  - \frac{1}{4\,}V^{(0)}\Phi_{\!\ssH}(\cph,\cph) 
  \Bigr],
\eq
etc. 
One last example of a basic fact: Nielsen identities give the structure
of the gauge parameter dependent vertex and self-energy order-by-order in perturbation
theory. It is important to stress at this point that the renormalized mass should
be replaced consistently with the use of \eqn{CPmren}.

To summarize, only $\cph$ is a meaningful quantity and a definition of
the real mass or of the total width is conventional. From \eqn{eqCP} one has
\bq
\srph\,\lgh = \Imb\,\Sigma_{\ssHH}(\cph), 
\eq
and it should be  evident, from \eqn{GPO}, that $\lgh \not= \sum_f\,
\Gamma\lpar H_c \to f\rpar$. The reason can be understood when we consider a 
simple example, a toy model with ${\cal L}_{\rm int}=m^2\,\phi\,\sigma^+\,\sigma^-$ 
(with massless $\sigma\,$-particles). Already at one-loop, we find
\bq
\Imb\,\Sigma_{\phi\phi}(s) = \frac{m^2}{16\,\pi^2}\,\pi,
\qquad 
\Imb\,\Sigma_{\phi\phi}(s_{\phi}) = \frac{m^2}{16\,\pi^2}\,\pi\,
\Bigl( 1 + \frac{1}{\pi}\,\atan{\frac{\gamma_{\phi}}{\mu_{\phi}}} \Bigr).
\label{cut}
\eq
While the first relation in \eqn{cut} (real $s$) satisfies the cutting 
equation~\cite{Cutkosky:1960sp} the second (complex $s$) does not. 
For a proper perspective it must be recalled that
when we expand, $\Sigma_{\ssHH}(\cph) = \Sigma_{\ssHH}(\rph) + \dots$, 
the cutting equation is restored at NLO but it will still be violated at
NNLO, as pointed out in Ref.~\cite{Grassi:2001bz}.
Therefore, our conventional definition of the Higgs total decay width  
will be $\Gamma_{\rm tot}(H_c) = \sum_f\,\Gamma\lpar H_c \to f\rpar$.

\begin{figure}[h]
\vspace{0.3cm}
\begin{picture}(140,30)(-120,0)
 \SetScale{0.5}
 \SetWidth{1.8}
 \GCirc(50.,0.){10.}{0} 
 \Gluon(10.,40.)(50.,0.){2}{7}
 \Gluon(10.,-40.)(50.,0.){2}{7}
 \DashLine(50.,0.)(90.,0.){3}
 \Line(100.,20)(100.,-20)
 \Line(110.,20)(110.,-20)
 \DashLine(120.,0.)(160.,0.){3} 
 \GCirc(160.,0.){10.}{0} 
 \Line(160.,0.)(200.,40.)
 \Line(160.,0.)(200.,-40.)
 \Line(210.,60.)(210.,20.)
 \Line(220.,60.)(220.,20.)
 \Line(210.,-60.)(210.,-20.)
 \Line(220.,-60.)(220.,-20.)
 \Line(230.,40.)(270.,40.)
 \Line(230.,-40.)(270.,-40.)
 \GCirc(270.,40.){10.}{0} 
 \GCirc(270.,-40.){10.}{0} 
 \ArrowLine(310.,60.)(270.,40.)
 \ArrowLine(270.,40.)(310.,20.)
 \ArrowLine(310.,-60.)(270.,-40.)
 \ArrowLine(270.,-40.)(310.,-20.)
 \Text(55,-30)[cb]{\Large $s_{\ssH}$}
 \Text(110,-50)[cb]{\Large $s_{\ssW}$}
\end{picture}
\vspace{1.5cm}
\caption[]{Gauge-invariant breakdown of the triply-resonant $g g \to 4\,$f 
signal into $gg \to H$ production, $H \to W^+ W^-$ decay and subsequent
$W \to {\bar f} f$ decays.}
\label{Multi}
\end{figure}
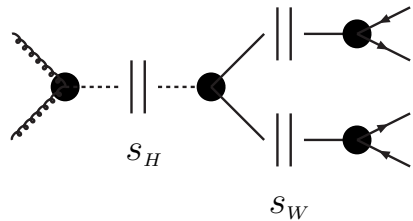

To set the stage, it may be well to recall that the breakdown of a process into 
products of pseudo-observables can be generalized to include unstable particles 
in the final state; an example is given in \fig{Multi} where the 
(triply-resonant) signal in $g g \to 4\,$f is split into a chain $gg \to H$ 
(production), $H \to W^+ W^-$ (decay) and $W \to {\bar f} f$ (decays).
\section{Pseudo-observables, on-shell observables and unitarity 
\label{UNI}}
When we consider all the possible decay channels of an {\em on-shell} standard
model Higgs boson we obtain that up to an on-shell mass $m_{\ssH} \approx 
140\,$GeV the Higgs boson is very narrow while the width rapidly increases 
after the opening of the $WW$ and $ZZ$ channels.

Even this statement should be carefully examined since $W$ and $Z$ bosons are
unstable particles to be removed from the in/out bases of the Hilbert space.
For real $W,Z$ masses the Higgs boson width is related to the cuts of the 
self-energy and the statement under examination is based on the (say one-loop) 
two-fermion cut, two-boson cut, etc.

Unitarity follows if we add all possible ways in which a diagram with
given topology can be cut in two separating $S$ from $S^{\dagger}$.  For
a stable particle the cut line, proportional to the positive energy part
of the propagator, contains a pole term
$2\,i\,\pi\,\theta(p_0)\,\delta(p^2+m^2)$, whereas there is no such
contribution for an unstable particle. We express $\Imb\,\Sigma$ in
terms of cut self-energy diagrams and repeat the procedure ad libidum,
therefore proving that cut unstable lines are left with no contribution,
i.e. unstable particles contribute to the unitarity of the $S-$matrix
via their stable decay products~\cite{Veltman:1963th}.

From this point of view the second cut of the Higgs self-energy (after the 
two-fermion cut) is the four-fermion cut, not the two-boson one (once again, 
the cutting of a line corresponding to an unstable particle contains no pole 
term). 
How bad is the choice of cutting two, {\em stable}, $W$ boson lines with respect to 
cutting four fermion lines and summing over all fermions, i.e. how bad is the 
on-shell approach, at least from a numerical point of view?

If one evaluates the ratio
\bq
\Gamma \lpar H \to VV \rpar \, \hbox{BR}\lpar V \to 2f\rpar \, 
\hbox{BR}\lpar V \to 2f'\rpar \; / \;  \Gamma \lpar H \to 2f + 2f' \rpar
\eq
the results of Ref.~\cite{Bredenstein:2006rh} show that the on-shell phase 
space for the $WW$ or $ZZ$ final state introduces an error of the order of 
$10\%$ near the threshold, which is still satisfactory. Using the complex 
mass scheme which, in turns violates unitarity, will improve upon the 
on-shell result since the internal $V$ masses are themselves complex poles.
Remarkably, the complex mass scheme represents a method which is, at the same 
time, predictive and gives the best available approximation to the use of 
a full (Schwinger-Dyson) re-summed theory, a formal solution of the 
problem which, however, poses an insurmountable barrier for the technology 
of today.
\section{Loop integrals with complex masses and invariants 
\label{allcmplx}}
In this section we analyze the correct definition of Feynman integrals
with complex masses and Mandelstam invariants. 
On a more formal bases one should say that unstable states lie in a natural
extension of the usual Hilbert space that corresponds to the second sheet of
the $S\,$-matrix; these states have zero norm and, therefore, escape the
usual prohibition of having an hermitian Hamiltonian with complex 
energy~\cite{Weldon:1975gu}.
On a more pragmatic level we use the guiding principle that Green's functions 
involving unstable particles should smoothly approach the value for stable 
ones (the usual Feynman $-\,i\,0$ prescription) when the couplings of the theory 
tend to zero. 

The whole problem can be summarized as follows: in the limit of zero
couplings all particles are stable and we define Green's functions in the
cut $s\,$-plane, where $s$ is the selected invariant to be continued into 
the complex plane. 
For the {\em free} theory of stable particles, according to Feynman 
prescription, the value of the argument of some function lies, say below 
the cut (which coincides for example to the real negative axis);
during continuation of $s$ we may cross the cut, which means that we have 
to continue the function into its second branch. 

For the simple case that we have just described the Green's function is then 
defined through its value on the principal branch in all quadrants but the 
second, where continuation to the second branch is required. 
This {\em new} function will have a cut on the positive imaginary axis and 
special problems may occur, especially when we want to do analytical 
continuation at the level of integrands and also internal masses in a 
given Feynman diagram are complex, as required by any realistic 
complex-mass scheme.
Green's functions are given in terms of Feynman parametric integrals and 
our main point will be: how to define the same integrals but properly 
continued to complex internal masses and complex external invariants? 
One of the difficulties of the problem lies in having masses and invariants 
complex at the same time which introduces subtleties in the analytical 
continuation which are not present if, say only masses or only invariants 
are made complex. 
\subsection{General setup \label{GS}}
To start our analysis, consider a scalar $\phi\,\sigma^2$ theory with 
$M_{\phi} > 2\,m_{\sigma}$, i.e. $\phi$ is unstable; the $\phi$ propagator 
(with $s = -\,p^2$) is
\bq
\Delta = \Bigl[ s - M^2_{\phi} + \Sigma_{\phi\phi}(s)\Bigr]^{-1},
\eq
where factors $(2\,\pi)^4\,i$ have been omitted.
The inverse function, $\Delta^{-1}(s)$ is analytic in the entire $s\,$-plane
except for a cut from $s = 4\,m^2_{\sigma}$ to infinity along the real
axis. The function is defined above the cut, $\Delta^{-1}(s + i\,0)$ and
the analytical continuation downwards is to the second Riemann sheet, i.e.
\bq
\Delta^{-1}_2(s - i\,0) = \Delta^{-1}(s + i\,0) = \Delta^{-1}(s - i\,0) + 
2\,i\,\pi\,\rho(s),
\label{AC}
\eq
where $2\,i\,\pi\,\rho(s)$ is the discontinuity of the function across the cut.
For a complete discussion of the analytical continuation see, e.g., 
Ref.~\cite{Brown:1992db}.

We need a few definitions which will help the understanding of the
procedure for the analytical continuation of functions defined through a
parametric integral representation. The logarithm is defined by
\bq
\ln^{(k)} z = ln^{(0)} z + 2\,i\,\pi\,k, 
\quad 
k = 0\,,\,\pm 1\,,\,\pm 2\,,\dots
\eq
where $\ln^{(0)} z$ denotes the principal branch ($ - \pi < \arg(z) \le + \pi$).
From now on we will omit the superscript that denotes the principal branch of
the logarithm. Let $z_{\pm} = z_0 \pm i\,0$ and $z= z_{\ssR} + i\,z_{\ssI}$, 
we define
\bq
\ln^{\pm} \lpar z\,;\,z_{\pm} \rpar = 
\Bigl\{
\begin{array}{ll}
{} \quad & \quad 
\ln z \pm 2\,i\,\pi\,\theta \lpar - z_0 \rpar\,\theta \lpar \mp z_{\ssI} \rpar \\
{} \quad & \quad 
\ln z \pm 2\,i\,\pi\,\theta \lpar - z_{\ssR} \rpar\,\theta \lpar \mp z_{\ssI} \rpar,
\end{array}
\label{variants}
\eq
i.e. the first Riemann sheet for all quadrants but the second where the 
function is defined in the second Riemann sheet. 

Our first definition of the $\ln^{\pm}\,$-functions in \eqn{variants} is the 
most natural in defining analytical continuation of Feynman integrals with a 
smooth limit into the theory of stable particles; the reason is simple, in 
case some of the particles are taken to be unstable we have to perform 
analytical continuation only when the corresponding Feynman diagram, in the 
limit of all (internal) stable particles, develops an imaginary part (e.g. above some 
normal threshold).
However, in all cases where the analytical expression for the diagram is known,
one can easily see that the result does not change when replacing $z_0$ with
$z_{\ssR}$, the second variant in \eqn{variants}. 

As we are going to discuss in the following sections there are cases where one 
would like to perform an analytical continuation at the level of integrand in the 
Feynman parametric representation of a given diagram; often the integration 
contour has to be distorted into the complex plane with the consequence that 
$z_{\ssR} \not= z_0$ and ${\rm sign}(z_{\ssR}) \not= {\rm sign}(z_0)$.
In this case we need a more general definition of $\ln^\pm$:
\begin{description}
\item[{\bf Definition:}] 
Let $z(\Gamma) \in C$ ($\Gamma \in R$) be an arbitrary complex function 
of $\Gamma$; when we want to continue $z_0 = z(0)$ (not in the second 
quadrant) to $z_f = z(\Gamma_f)$ we must look for a real $\Gamma_c$ with 
$0 < \Gamma_c < \Gamma_f$ such that $z_c = z(\Gamma_c)$ is real and negative 
(for simplicity we assume the case a monotonic $z_{\Gamma} = z(\Gamma)$): 
then, $\forall \Gamma\,:\, \Gamma \ge \Gamma_c$ we replace $\ln z$ with 
its analytical continuation into the second Riemann sheet,
\bq
\ln^{\pm} \lpar z_{\Gamma}\,;\,z_0 \rpar = 
\ln z_{\Gamma} \pm 2\,i\,\pi\,\theta \lpar - \Reb\,z_{\Gamma} \rpar\,
\theta \lpar \mp \Imb\,z_{\Gamma} \rpar.
\label{cdef}
\eq
For all practical purposes \eqn{cdef} can be replaced with the second variant
of \eqn{variants} (with $z \to z_\Gamma$) which, from now on, will be our definition
of analytical continuation.
\end{description}
\subsection{Analytical continuation of the Euler dilogarithm \label{ED}}
We consider now the Euler's dilogarithm, $\li{2}{z}$; if we denote its
principal branch by $\mbox{Li}^{(0,0)}_2(z)$ ($0 < \arg(z-1) < 2\,\pi$), than for 
any branch (see, e.g.~\cite{HTF})  we have
\bq
\mbox{Li}^{(n,m)}_2(z) = \mbox{Li}^{(0,0)}_2(z) + 2\,n\,\pi\,i\,\ln^{(0)} z + 4\,m\,\pi^2,
\quad
n\,,\,m = 0\,,\,\pm 1\,,\,\pm 2\,,\dots
\label{ACLi}
\eq
The question that we want to analyze is the following: given
\bq
\mbox{Li}_2(M^2+ i\,0) = 
- \int_0^1\,\frac{dx}{x}\,\ln \lpar 1- M^2\,x - i\,0\rpar,
\qquad
\Imb\,\mbox{Li}_2\lpar M^2 + i\,0\rpar = \pi\,\ln M^2\,\theta\lpar M^2 - 1\rpar,
\eq
how do we understand \eqn{ACLi} in terms of an integral representation?
Let us consider the analytical continuation from $z^+ = M^2 + i\,0$ to $z = M^2 - 
i\,M\,\Gamma$ and define
\bq
I = - \int_0^1\,\frac{dx}{x}\,\ln^- \lpar 1 - z\,x\,;\, 1 - z^+\,x\rpar.
\eq
With $\chi(x) = 1 - z\,x = 1 - (M^2-i\,M\Gamma)\,x$, we have $\chi(0) = 1$ and
$\chi(1)= \lpar 1 - M^2\,,\,M\,\Gamma\rpar$. If $M^2 > 1$ we have that $\chi$
crosses the positive imaginary axis with $\Imb\,\chi = \Gamma/M$. As a
result we obtain
\bq
I = \mbox{Li}^{(0,0)}_2(z) + 2\,i\,\pi\,\ln M^2,
\eq
which is not the expected result since $I$ does not reproduce the correct 
continuation of $\mbox{Li}_2$ given in \eqn{ACLi}. 
The mismatch can be understood by observing that $\ln^-$ has a cut along 
the positive imaginary axis (of $\chi$) and, in the process of continuation, 
with $x \in [0,1]$, we have been crossing the cut.
Nevertheless, we insist on defining analytical continuation at the level of
integrand, instead of working directly on the result, because it is the only
practical way of dealing with multi-loop diagrams where an exact result is not
known. 
The solution consists in deforming the integration contour, therefore
defining a new integral,
\bq
I_{\ssC} = 
\int_{\ssC}\,\frac{dx}{x}\,\ln^- \lpar 1 - z\,x\,;\, 1 - z^+\,x\rpar,
\eq
where the curve $C$ is given by two straight segments, 
$0 \le x \le 1/M^2 - \ep$ and $1/M^2 + \ep \le x \le 1$
($\ep \to 0^+$), plus a curve $C'$ defined by
\bq
C'(u)\; :\; \{ x = u + i\,\frac{1 - M^2\,u}{M\,\Gamma}\},
\quad  
\frac{1}{M^2 + \Gamma^2} \le u \le \frac{1}{M^2},
\eq
The integral over $C'$ is downwards on the first quadrant an upwards on the 
second (along the cut of $\ln^-$). Integration of $\ln^-$ over $C'$ gives 
$- 2\,i\,\pi\,( \ln M^2 - \ln z)$, showing that
\bq
\mbox{Li}^{(1,0)}_2(z)= I_{\ssC},
\eq
the correct analytical continuation. Therefore we can extend our integral, by
modifying the contour of integration, to reproduce the right analytical 
continuation of the dilogarithm.
\subsection{Continuation of analytical results \label{contan}}
Having introduced a simple example, we consider now one-loop two-point 
functions where both masses and the external invariant are made complex. 
Let
\bq
\chi(x) = s_{\ssP}\,x^2 + \lpar m^2_2 - m^2_1 - s_{\ssP} \rpar\,x + m^2_1,
\eq
\bq
s_{\ssP} = M^2 - i\,\Gamma\,M, \qquad
m^2_i = \mu^2_i - i\,\gamma_i\,\mu_i.
\eq
The function $B_0$ is originally defined, for real $s_{\ssP}$ and real 
(equal for simplicity) internal masses, by 
\bq
B_0\lpar M^2\,;\,\mu,\mu\rpar = \frac{1}{{\bar\ep}} - 
\intfx{x}\,\ln ( \chi - i\,0),
\label{B0integ}
\eq
where ${\bar\ep}^{-1}= 2/(4-n) - \gamma_{\ssE} - \ln\pi$ 
($\gamma_{\ssE} \approx 0.5772$ being the Euler-Mascheroni constant), and we 
need the analytical continuation to arbitrary values of $s_{\ssP}$ 
(i.e. $M^2 \to M^2 - i\,M\,\Gamma$ with $\Gamma > 0$); 
we assume, for a moment, real internal masses ($\gamma= 0$) and 
$M^2 > 4\,\mu^2$; the analytical result is
\bq
B_0\lpar M^2\,;\,\mu,\mu\rpar = 
\frac{1}{{\bar\ep}} - \ln\frac{\mu^2}{\mu^2_{\ssR}} 
+ 2 -\beta\,\ln\frac{\beta+1}{\beta-1},
\label{anB0}
\eq
where $\mu_{\ssR}$ is the renormalization scale and $\beta^2 = 1 - 4\,\mu^2/M^2$.
For the continuation to $M^2 \to M^2 - i\,M\,\Gamma$, 
$\mu^2 \to \mu^2 - i\,\gamma\,\mu$  we have to compute the logarithm of 
$z^{\UST} =z^{\UST}_{\ssR} + i\,z^{\UST}_{\ssI}$, which is a function of 
$\Gamma, \gamma$ (interacting theory of unstable particles). 
Let
\bq
z^{\ST}_{\pm} = 
\lim_{\Gamma, \gamma \to 0}\,z^{\UST}_{\ssR} \pm i\,0 =
z^{\ST}_{\ssR} \pm i\,0,
\eq
where the $\pm i\,0$ follows from Feynman prescription $\mu^2 \to \mu^2 - i\,0$. 
We use the second variant of \eqn{variants} and define
\bq
\ln^{\pm} \lpar z^{\UST}\,;\,z^{\ST}_{\pm} \rpar = 
\ln z^{\UST} \pm 2\,i\,\pi\,\theta \lpar - z^{\UST}_{\ssR} \rpar\,
\theta \lpar \mp z^{\UST}_{\ssI} \rpar,
\eq
which satisfies
\bq
\lim_{\Gamma, \gamma \to 0}\,\ln^{\pm} \lpar z^{\UST}\,;\,
z^{\ST}_{\pm} \rpar = \ln z^{\ST}_{\pm},
\eq
and it is equivalent to have $\ln\!z$ on the second Riemann sheet, but only 
when $z$ is continued into the second quadrant.
There is one awkward possibility; it corresponds to starting from $z^{\ST}_-$ with 
$z^{\ST}_{\ssR} < 0$ and requiring continuation to $z^{\UST}_{\ssI} > 0$ and 
$z^{\UST}_{\ssR} > 0$.  
Using \eqn{anB0} we derive
\bq
B_0\lpar M^2\,;\,m,m\rpar \to \frac{1}{{\bar\ep}} - 
\ln\frac{m^2}{\mu^2_{\ssR}} 
+ 2 -\beta_c\,\ln^- \lpar \frac{\beta_c+1}{\beta_c-1}\,;\,
\frac{\beta+1}{\beta-1}\rpar
\label{anCB0}
\eq
where $\beta^2_c = 1 - 4\,m^2/s_p$. 
It is worth noting that there is never a problem when internal masses are 
real and we continue to complex $p^2$. 
Otherwise we first continue to complex internal masses using the fact that 
internal (complex) squared masses have a negative imaginary part. 
Consider this continuation for one-loop diagrams: with $L\,$-external legs
we can always fix a parametrization where the coefficient of $m^2_1$ is
$1-x_1$, the one of $m^2_i$ is $x_{i-1}-x_i$, up to $m^2_{\ssL}$ which has
coefficient $x_{\ssL-1}$ where the parameters satisfy $0 \le x_{\ssL-1} \le
\,\dots\,\le x_1 \le 1$ (i.e. all coefficients are non-negative). Less
straightforwardly the same holds for multi-loop diagrams.

Then we continue from $\Gamma = 0$; considering \eqn{anCB0} and denoting by 
$\zeta$ the ratio $(\beta_c+1)/(\beta_c-1)$ we have $\Reb\,\beta^2_c >0$ (in 
the region above real thresholds) and $\Imb\,\beta^2_c > 0$ for $\Gamma= 0$. 
At $\Gamma = (M/\mu)\,\gamma$ $\beta^2_c$ crosses the positive real axis from 
above; this corresponds to $\zeta$ crossing the cut where we move into the 
second Riemann sheet of the logarithm of \eqn{anCB0}. 
After that one has $\Imb\,\zeta > 0$ and the forbidden region is 
reached when $\Reb\,\zeta > 0$, which corresponds to $\mid \beta_c \mid > 1$. 
Once again, for $\gamma = 0$ the latter is never satisfied. In general, the 
condition requires solving a cubic equation in $\Gamma$ with only one real, 
negative, solution. 
The forbidden region requires, therefore, $\Gamma < 0$.

To continue our analysis of one-loop functions, where analytical results are 
known, we only need to define
\bq
\mbox{Li}^{\mp}_2\lpar z^{\UST}\,;\,z^{\ST}_{\mp}\rpar = 
\li{2}{z} \mp 2\,i\,\pi\,\theta \lpar z^{\UST}_{\ssR} - 1 \rpar\,
\theta \lpar \pm z^{\UST}_{\ssI} \rpar\,\ln z^{\UST}.
\label{li2pm}
\eq
For our purposes, namely for the processes that we are considering, we only 
need one additional function. 
The most general scalar three-point function that is needed will be
\bq
C_0\lpar 0,0,p^2\,;\,m_1,m_2,m_3 \rpar = \frac{1}{p^2}\,\Bigl\{ \sum_{i=1,3}\,
(-1)^{\delta_{i3}}\,\Bigl[
\mbox{Li}_2\lpar\frac{x_0-1}{x_0-x_i}\rpar - \mbox{Li}_2\lpar\frac{x_0}{x_0-x_1}\rpar \Bigr]
+ \ln x_0\,\eta\lpar x_1-x_0\,,\,x_2-x_0\rpar\},
\label{etaAC}
\eq
with four different roots
\bq
x_0 = 1 + \frac{m^2_1-m^2_2}{p^2}, 
\qquad\quad
x_3 = \frac{m^2_3}{m^2_3-m^2_2},
\qquad\quad
x_{1,2} = 
\frac{p^2+m^2_1-m^2_3 \mp 
\lambda^{1/2}\lpar -p^2,m^2_1,m^2_3\rpar}{2\,p^2},
\eq
where $\lambda$ is the K\"allen function.
Analytical continuation requires the replacement $\mbox{Li}_2 \to \mbox{Li}^-_2$ with limiting
(free theory of stable particles) cases given by
\bq
x_1 \to x_1 -i\,0, 
\qquad\quad 
x_2 \to x_2 + i\,0,
\qquad\quad
x_3 \to x_3 - i\,{\rm sign}(m^2_1 - m^2_2)\,0.
\eq
As a final observation, there is no need to continue the square root 
$\beta_c$ in \eqn{anCB0} below threshold ($\beta^2_c < 0$) since in this 
case $\beta_c$ is imaginary and the change of sign when we move from
the principal root is compensated in the product $\beta_c$ times the logarithm.

Finally, in \eqn{etaAC} and for one-loop processes with more scales and more 
than three legs one has to introduce a generalization of 't Hooft-Veltman 
$\eta\,$-functions~\cite{'tHooft:1978xw} on the second Riemann sheet. The 
definition is as follows: 
\bq
\ln^- (x\,y) = \ln^- x + \ln^- y + \eta^-(x,y),
\eq
\bq
\eta^-(x,y) = 2\,i\,\pi\,\Bigl\{
            \theta(x_{\ssI})\,\Bigl[ \theta( - x_{\ssR})
          - \theta(y_{\ssI})\,\theta( - z_{\ssI})\Bigr]
          + \theta(y_{\ssI})\,\theta( - y_{\ssR})
          + \theta(z_{\ssI})\,\Bigl[ \theta( - x_\ssI)\,\theta( - y_{\ssI})
          - \theta( - z_{\ssR}) \Bigr]\Bigr\},
\eq
with $z= x\,y$ and $z = z_{\ssR} + i\,z_{\ssI}$ etc.
\subsection{Continuation at the integrand level \label{CIL}}
We now turn to analytical continuation at the integrand level, according 
to our procedure where all Feynman integrals are treated according to their 
parametric integral representation.

Let us consider the specific example of the previous section: a scalar 
two-point function corresponding to two internal equal masses, 
$m^2= \mu^2 - i\,\mu\,\gamma$ and $s_p = M^2 - i\,M\,\Gamma$. 
Due to the symmetry of $\chi(x)$, in \eqn{B0integ} the integral 
with $0 \le x \le 1$ can be written as twice the same integral with 
$0 \le x \le 1/2$; the argument of the logarithm goes from $\Reb\,\chi = \mu^2 > 0$ 
to $\Reb\,\chi = \mu^2 - M^2/4 < 0$ (above threshold) with 
$\Imb\,\chi = -\,i\,0$.
We have to define the analytical continuation $M^2 \to M^2 - i\,M\,\Gamma$;
since, for any $x$, $\chi$ cannot cross the cut, it must be analytically 
continued into a second Riemann sheet above the cut.
A similar situation occurs for complex internal masses: the integration
with respect to $x$ is
\bq
\chi = \lpar \mu^2\,,\,- i\,\mu\,\gamma \rpar \quad \to \quad 
\chi = \lpar \mu^2 - \frac{1}{4}\,M^2\,,\, - i\,\mu\,\gamma + 
\frac{i}{4}\,M\,\Gamma \rpar.
\eq
Let $X = x\,(1-x)$ with $0 \le X \le 1/4$, select a value for $x$, when
$\Gamma \geq (\mu\,\gamma)(M\,X)$ continuation is into the second Riemann 
sheet.
Of course, for $M^2 < 4\,\mu^2$, $\chi$ remains on the first Riemann sheet for
all values of $\Gamma$. The variable $\chi$ is such that
\bqa
\Reb\,\chi &=& 0 \qquad \hbox{for} \quad 
x = R_{\pm} = \frac{1}{2}\,\Bigl[ 1 \pm \sqrt{1 - 4\,\frac{\mu^2}{M^2}}\Bigr].
\nl
\Imb\,\chi &=& 0 \qquad \hbox{for} \quad 
x= I_{\pm} = \frac{1}{2}\,\Bigl[ 1 \pm
\sqrt{1 - 4\,\frac{\mu \gamma}{M \Gamma}}\Bigr].
\label{RpmIpm}
\eqa
The second equation requires $M \Gamma \ge 4\,\mu \gamma$ for $I_{\pm}$ 
to be real and $\in\,[0\,,\,1]$. 
At $x = I_{\pm}$ the condition $\Reb\,\chi \le 0$ requires 
$\Gamma \mu \le M \gamma$. 
Therefore, for those values of $\Gamma$ and $x$ that satisfy the conditions
\bq
4\,\frac{\mu}{M}\,\gamma \le \Gamma \le \frac{M}{\mu}\,\gamma,
\qquad\qquad
I_-\leq x \leq I_+,
\qquad\qquad
R_-\leq x \leq R_+,
\label{xcrossing}
\eq
we have $\Reb\chi \leq 0$, $\Imb\chi\geq 0$ and $\ln\chi$ must be continued 
into the second Riemann sheet. 

\begin{figure}[th]
\begin{center}
\includegraphics[bb=152 441 476 706,width=7cm]{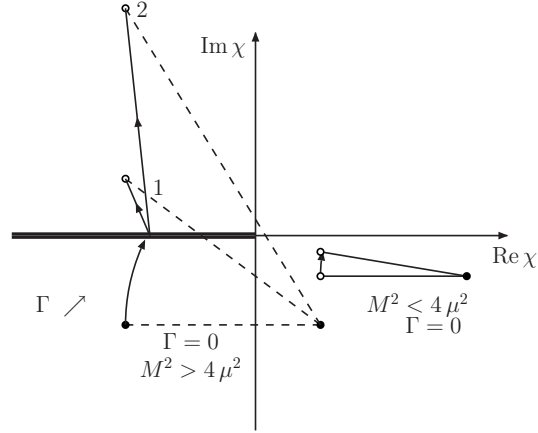}
\end{center}
\vspace{-0.6cm}
\caption[]{\label{chip}
Analytical continuation from real $p^2$ to complex $p^2$ as seen in the 
$\chi\,$-plane with $\chi(x)= - s_{\ssP}\,x\,(1-x) +\mu^2 - i\,\mu\,\gamma$, 
with $s_{\ssP} = M^2 - i\,M\,\Gamma$ and $x\in [0,1]$. 
Solid lines represent the continuation for a low value of $M$ with a very small 
value for $\Gamma$. With increasing values for $M$ we reach the situation 
illustrated by the dot-lines, $\chi$ moving into the second quadrant, i.e. 
$\chi$ on the second Riemann sheet. Case $1$ holds for
$\Gamma < (M/\mu)\,\gamma$ whereas case $2$ holds for
$\Gamma > (M/\mu)\,\gamma$. 
Black circles correspond to $x=0, x=1$ whereas white circles correspond to 
$x= 1/2$.
}
\end{figure}

The new definition of the $B_0\,$-function is as follows:
\bq
B_0 = \frac{1}{{\bar\ep}} - \intfx{x}\,\ln^{-}\lpar \chi\,;\,\chi_- \rpar,
\qquad\quad
\chi_- = \chi\bmid_{\Gamma,\gamma_i = 0} - i\,0,
\label{safe}
\eq
Different possibilities are illustrated in \fig{chip} where we show $\chi(x)$
for two equal (complex) internal masses. 
\\
In any realistic application the
complex pole equation returns, for low values of $M$, small values of
$\Gamma$ and $\Reb\,\chi$ is always positive, never requiring analytical 
continuation into another sheet; when $M$ increases $\Gamma$ increases too and
we find values of $\chi$ that requires the continuation $\ln \to \ln^-$. This 
will happen for $x \ge I_-$ in case $1$ (which requires 
$\Gamma < (M/\mu)\,\gamma$) and for $x > R_-$ in case $2$ (which requires 
$\Gamma > (M/\mu)\,\gamma$).

The same example can be discussed in the $x$ complex plane; in this case, when
$M > 2\,\mu$ and $\Gamma= \gamma= 0$, the cut is on the real axis between $R_-$ 
and $R_+$ (\eqn{RpmIpm}) and the integration is $0 < x < 1/2$. The integral is 
originally defined ($\Gamma, \gamma = 0$) above the cut, i.e. for $x + i\,0$. 
Analytical continuation means that for increasing imaginary parts we reach a 
point where the integration path is continued into the second Riemann sheet 
(at $\Gamma = (M/\mu)\,\gamma$, we have $I_- = R_-$ and, for higher values of 
$\Gamma$, the continuation to the second Riemann sheet is required as soon as 
the cut is reached).
From this point of view the integral is better understood in terms of
a variable $z(x)= u + i\,v$, such that $\chi = - M^2\,z\,(1-z) + \mu^2$ and the
integration is performed along the curve
\bq
v\,(1 \!-\! 2\,u) = \frac{Z}{M^2},
\qquad
Z = \mu\,\gamma - M\Gamma\,x\,(1-x),
\qquad
u = \frac{1\!-\!U}{2},
\qquad
U^4 + \Big[ 4\,x(1\!-\!x) - 1\Big]\,U^2 - 4\,\frac{Z^2}{M^4} = 0.
\label{ztransf}
\eq
Note that $x= I_-$ corresponds to $z= I_-$, real. For real internal masses we
have $z(0)= 0$ and $z(1/2)= (\Gamma/(2 M))^{1/2}\,(1 - i/2)$.
In the $z\,$-plane the logarithm has a cut on the positive real axis between
$R_-$ and $R_+$.
 
\begin{figure}[ht!]
\begin{center}
\includegraphics[bb=274 486 485 706,width=5cm]{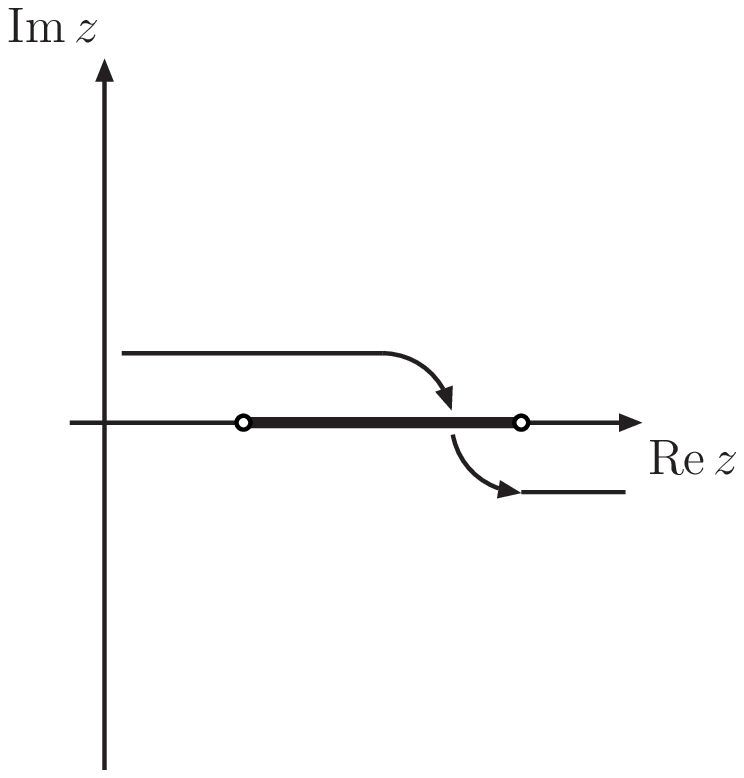}
\hspace{3cm}
\includegraphics[bb=274 486 485 706,width=5cm]{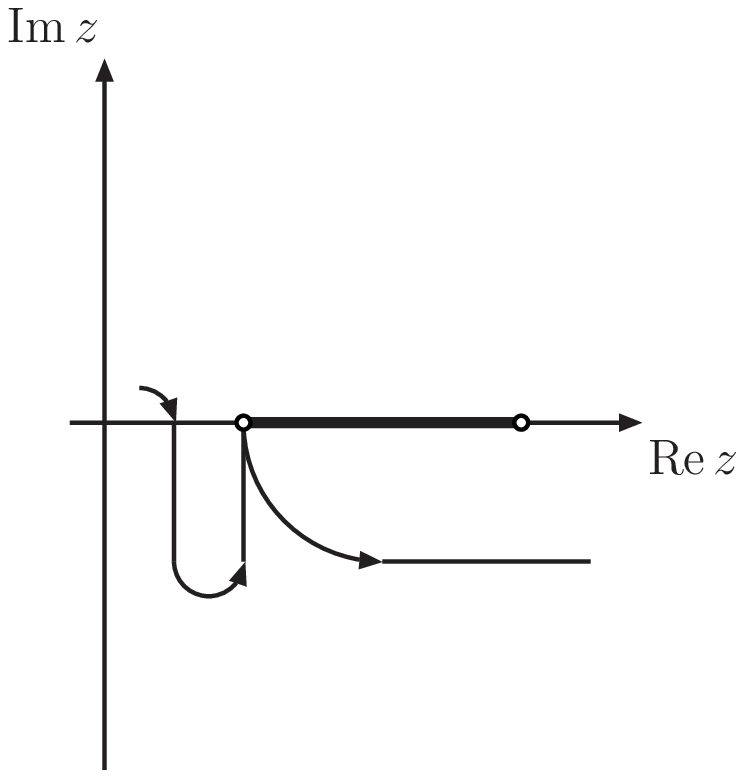}
\end{center}
\vspace{-0.4cm}
\caption[]{\label{xplane}Analytical continuation of a $B_0\,$-function
as seen in the $z\,$-plane with a cut along the positive real axis
between $R_-$ and $R_+$ (\eqn{RpmIpm}). In the first part the
integration path reaches the point $I_-$ (\eqn{RpmIpm}) and continuation
after $z = I_-$ is in the second Riemann sheet.  In the second part,
where $I_- < R_-$ continuation must be, once again, in the second
Riemann sheet; therefore the integration path which has moved into the
lower half-plane must be deformed to cross the cut before moving once
more into the lower half-plane (but on the second Riemann sheet).}
\end{figure}

It is worth mentioning that case $2$ of \fig{chip} corresponds to an 
integration path that crosses the cut of $\ln^-$ across the positive imaginary 
$x\,$-axis, similar to the case of the dilogarithm discussed above. 
Therefore, the correct analytical continuation, for case $2$, goes as follows: 
the integration path in $z\,$-space (\eqn{ztransf}) is moved into the complex 
plane and goes into the lower half-plane instead of reaching the cut of the
logarithm (which is between $R_-$ and $R_+$, see also \fig{xplane}). 

In order to insure that the analytically continued integral has a smooth limit 
$\Gamma, \gamma \to 0$ we deform the integration path by insisting that the 
cut (of $\ln$) must be crossed at $z= R_-$ (note that for case $2$ we have 
$I_- < R_-$) where we perform a continuation into the second Riemann sheet. 
In this way we add to $B_0$ (on top of a factor $-\,2\,i\,\pi\,\beta$) a new 
contribution which is easily computed in the $x\,$-plane and it is related 
to the discontinuity of $\ln\chi$ along a curve $C$ parametrized by  
\bq
C(t)\; :\; \{ x = \frac{1-t}{2} + i\,f(t)\}, \qquad
f(t) = \frac{1}{2}\,\Bigl\{ -\,\frac{\Gamma}{M}\,t + \Bigl[ \lpar 1 + 
\frac{\Gamma^2}{M^2} \rpar\,t^2 - \beta^2 \Bigr]^{1/2} \Bigr\},
\eq
with $\beta^2 = 1 - 4\,\mu^2/M^2 > 0$ and where $\bar{\beta} < t < \beta$;
here $\bar{\beta}$ is the value of $t$ where $\Reb\,\chi(t) = \Imb\chi(t) = 0$.
The integral over $C$ is on the segment $\Reb\,\chi = \pm \ep$ with 
$\ep \to 0^+$, from $\mu^2 \Gamma/M - \mu \gamma > \Imb\chi > 0$ on the first 
Riemann sheet and from $\mu^2 \Gamma/M - \mu \gamma < \Imb\chi < 0$ on the 
second Riemann sheet.
Therefore, we have to add to $B_0$ an additional term $-\,2\,i\,\pi\,\Delta\,A$ 
with
\bq
\Delta A=  A(\bar{\beta}) - A(\beta),
\qquad
A(t) = \lpar 1 + i\,\frac{\Gamma}{M} \rpar\,t - i\,  
\Bigl[ \lpar 1 + \frac{\Gamma^2}{M^2} \rpar\,t^2 - \beta^2 \Bigr]^{1/2}.
\eq
Note that in the limit $\Gamma, \gamma \to 0$ we have $\bar{\beta} \to \beta$ 
and this additional term vanishes. 
Furthermore, $A(\beta)= \beta$ and $A(\bar{\beta}) = \beta_c$, with 
$\beta^2_c = 1 - 4\,m^2/s_p$. 
Therefore, we reproduce the correct result of \eqn{anCB0}. 
The recipe is, therefore, replace $\ln$ with $\ln^-$ in the integrand but 
deform the integration contour in order to avoid crossing of the positive 
imaginary $\chi\,$-axis when this would occur.

In summary, our result with a simple example we observe that
\bq
\mbox{Li}_n \stackrel{\rm {\tiny{Analyt. Cont.}}}{\longmapsto} \mbox{Li}^-_n,
\quad
\mbox{Li}^-_{n+1}(z) \not= \int_0^z\,\frac{dx}{x}\,\mbox{Li}^-_n(x),
\eq
since deformation of the integration contour is required for the general case.
\subsection{Narrow width approximation \label{NWA}}
The practical implementation for higher point (or higher loop) functions 
presents a formidable technical problem, due to the higher dimension of the 
$x\,$-space; more will be explained in \sect{ACCD} but, for this reason, we have 
also considered analytical continuation in narrow-width-approximation 
(hereafter NWA). 
Here we replace $\ln$ with $\ln^-$ (or $\ln^+$) at the integrand level and 
do not perform any deformation of the integration (hyper-)contour. 
The resulting expression is expect to have a range of validity given by 
$\Gamma \ll M$. 
Numerical investigation of the Higgs complex pole shows that NWA returns 
reliable results when compared with the exact expression. 
The rationale for analytical continuation in NWA is based on the fact that, 
as we are going to show, all higher-point (higher-loop) functions admit 
integral representations with integrand of logarithmic nature (one-loop) 
or, at most, of poly-logarithmic nature (multi-loop).

Consider now the extension to complex variables of an arbitrary scalar 
three-point function $C_0$ (in NWA), defined by
\bq
C_0 = \int_{\scriptstyle 0}^{\scriptstyle 1}\,dx_1\,
\int_{\scriptstyle 0}^{\scriptstyle x_1}\,dx_2\,V^{-1-\ep/2}(x_1,x_2),
\label{eqdefC0}
\eq
where $n= 4 - \ep$ and $V$ is a quadratic form
\bq
V(x_1,x_2) = a\,x_1^2 + b\,x_2^2 + c\,x_1\,x_2 + d\,x_1 +
e\,x_2 + f - i\,0 \equiv x^t\,H\,x + 2\,K^t\,x + L ,
\eq
whose coefficients are related to the internal masses and the external
momenta by the relations $H_{ij} = -\,\spro{p_i}{p_j}, L = m^2_1$ and
\bqa
K_1 &=& \frac{1}{2} \, ( \spro{p_1}{p_1} + m_2^2 - m_1^2 ),
\quad
K_2 = \frac{1}{2} \, ( \spro{P}{P} - \spro{p_1}{p_1} + m_3^2 - m_2^2 ),
\label{Konetwo}
\eqa
with $P = p_1 + p_2$. 
Let us define the usual Bernstein - Sato - Tkachov (hereafter BST) factors 
(see Ref.~\cite{Ferroglia:2002mz}) as 
$B_3 = L - K^t\,H^{-1}\,K$ and BST co-factors $X = -\,H^{-1}\,K$.
It is convenient to introduce special notations, $X_0 = 1, 
\, X_3 = 0$, and $V(\widehat {i\;i+1})$ to denote contractions, i.e.
\bq
V(\widehat{0\;1}) = V(1,x_1), \quad
V(\widehat{1\;2}) = V(x_1,x_1), \quad 
V(\widehat{2\;3}) = V(x_1,0).
\eq
In this way we obtain a simple integral representation
\bq
C_0 = \frac{1}{B_3}\,\Bigl\{ \frac{1}{2} +
\int_0^1\,dx_1\,\Bigl[ \int_0^{x_1}\,dx_2\,
\ln\,V(x_1,x_2) - 
\frac{1}{2}\,\sum_{i=0}^{2}\,(X_i - X_{i+1})\,\ln V(\widehat{i\;i+1}).
\Bigr]\Bigr\}.
\label{sintr}
\eq
When some or all the invariants are complex, $P^2= -\,s_{\ssP}$ with
$s_{\ssP} = M^2 - i\,\Gamma\,M$ and $m^2_i = \mu^2_i - i\,\gamma_i\,\mu_i$
(in realistic cases, e.g. decay of an unstable particle, $p^2_{1,2}$ are real)
we define
\bq
V_- = V\bmid_{\Gamma,\gamma_i = 0},
\eq
which includes the $-\,i\,0$ prescription and write
\bqa
C_0 &=& \frac{1}{B_3}\,\Bigl\{ \frac{1}{2} +
\int_0^1\,dx_1\,\Bigl[ \int_0^{x_1}\,dx_2\,
\ln^{-}\lpar V(x_1,x_2)\,;\,V_-(x_1,x_2)\rpar 
\nl
{}&-& \frac{1}{2}\,\sum_{i=0}^{2}\,(X_i - X_{i+1})\,
\ln^{-}\lpar V(\widehat{i\;i+1})\,;\,V_-(\widehat{i\;i+1})\rpar
\Bigr]\Bigr\}.
\label{sintrC}
\eqa
For instance, with $P^2= - M^2 + i\,M\,\Gamma$, $p^2_{1,2}= 0$ and
$m_{1,3} = 0$, $m^2_2 = \mu^2 - i\,\mu\,\gamma$ we find that when
$\Gamma/\gamma \le \mu/M$ $\ln V$ must be continued to the second 
Riemann sheet for $0 \le x_2 \le (\mu\gamma)/(M\Gamma)$.

Starting with an integral representation of a three-point function where the 
integrand is the logarithm of a polynomial in parametric space is the safest 
way of performing analytical continuation; of course, going beyond NWA 
requires contour deformation but even the latter admits a consistent numerical 
implementation. 
Nor should one fail to notice 't Hooft and Veltman emphasis, in their seminal 
work~\cite{'tHooft:1978xw}, on this subject: they put up warning signs about 
continuation of their result to complex momenta.

For higher point functions ($L = D,E,F,\,\dots$) we apply the BST 
algorithm~\cite{Passarino:2001wv} as many times as it is needed to produce 
logarithms in the integrand and proceed by replacing $\ln$ with $\ln^-$,
\bq
L = \int_{\{x\}} d\{x\}\,\ln \lpar \chi(\{x\}) - i\,0\rpar \to 
    \int_{\{x\}_{C(\chi)}} d\{x\}\,\ln^-\,\chi(\{x\}),
\eq
where $\{x\}$ is the $x_1,\,\dots\,,x_n$ simplex and $\{x\}_{C(\chi)}$ is 
the path that avoids crossing the positive imaginary $\chi\,$-axis. 
NWA amounts to the identification $\{x\}_{C(\chi)} \equiv \{x\}$. 

For multi-loop integrals other functions must be extended, e.g. we will 
use \eqn{li2pm}, with similar results for all generalized Nielsen 
polylogarithms~\cite{Kolbig:1983qt}
\bq
S^-_{n,p}(z^{\!\UST}\!;\!z^{\ST}_-) =\,
  S_{n,p}(z^{\!\UST})\, 
+ \,\sum_{k=1}^p\frac{(- 2i\pi)^{\!k}}{k\,!}\bigg[ 
    S_{n,p-k}(z^{\!\UST})\,
  - \sum_{j=0}^{n-1}\frac{\ln^j\!z^{\!\UST}\!}{j\,!}
    S_{n-j,p-k}(z^{\!\UST})
  \bigg]
  \theta(z^{\!\UST}_{\ssR} \!-\! 1 )\,\theta(z^{\!\UST}_{\ssI}),
\eq
which is derived by using $\ln^+$ in the integral representation of the 
generalized Nielsen polylogarithms.
Since it has been shown that multi-loop diagrams can be written as integrals of 
multivariate generalized Nielsen polylogarithms~\cite{Uccirati:2004vy} our recipe
gives the analytical continuation to all orders in perturbation theory,
but one does have to be careful in one respect: using familiar relations such
as splitting of logarithms should be done with a grain of salt.
\subsection{Analytical continuation and contour deformation \label{ACCD}}
Exact analytical continuation at the integrand level can be performed by
deforming the integration contour into the complex parametric space
(for a general treatment see Ref.~\cite{Nagy:2006xy}). In this 
case we need the general definition of $\ln^-$ given in \eqn{cdef}. 

\begin{figure}[ht!]
\begin{center} 
\includegraphics[bb=69 252 505 681,width=9cm]{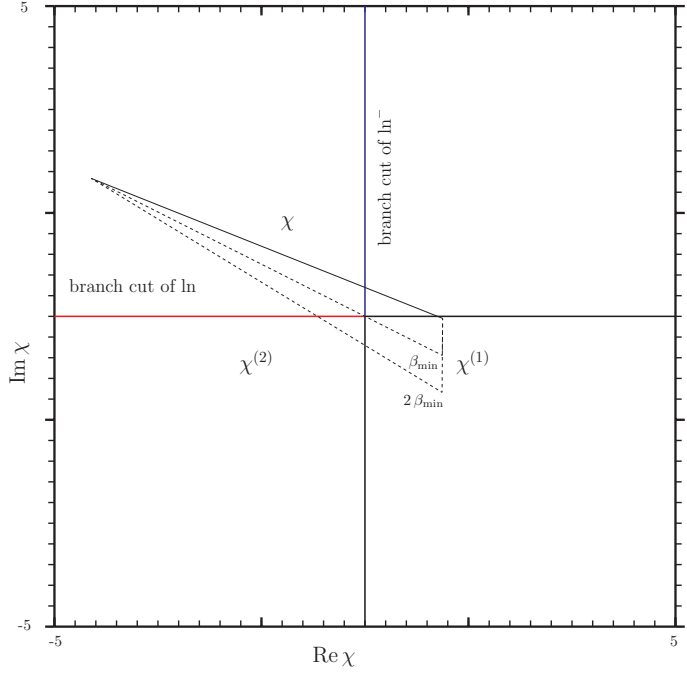}
\end{center}
\vspace{-0.5cm}
\caption[]{\label{c_def}Example of contour deformations in computing a
scalar two-point functions with equal (complex) internal masses and
complex $p^2$.}
\end{figure}

To illustrate contour deformation we consider, once again, the case of a 
$B_0\,$-function with equal (complex) internal masses. If
\bq
M > 2\,\mu
\qquad
\mu\,\Gamma - M\,\gamma > 0,
\eq
the function $\chi(x), x \in [0,\frac{1}{2}]$ crosses the positive imaginary axis
(the branch cut of $\ln^-$). To avoid crossing we deform the $x\,$-integration
into
\bqa
1) \quad &{}& \quad x = i\,\frac{\Gamma}{M}\,\beta\,t,
\nl
2) \quad &{}& \quad x = \frac{1}{2}\,t + i\,\frac{\Gamma}{M}\,\beta\,(1 - t),
\eqa 
with $t \in [0,1]$ and $\beta$ a free parameter. 
For $\chi^{(1)}$ we require $\Imb\,\chi^{(1)}(t) < 0, \forall t \in[0,1]$; 
this is possible if $\beta < \beta_{\rm max}$, with
\bq
\beta_{\rm max} = 
\frac{1}{2}\,\frac{M^2}{\Gamma^2}\,\Bigl[ 
1 + \sqrt{1 + 4\,\frac{\mu\,\gamma\,\Gamma}{M^3}}\Bigr].
\eq
For $\chi^{(2)}$ we require that $\Reb\,\chi^{(2)}(t) = 0$ corresponds to
$\Imb\,\chi^{(2)}(t) < 0$, which requires $\beta > \beta_{\rm min}$,
where $\beta_{\rm min}$ is the largest, real, solution of
the following equation
\bq
\mu\,\lpar \frac{\Gamma}{M^2} - \frac{1}{4\,\Gamma} \rpar\,
\lpar \Gamma\,\mu - M\,\gamma \rpar\,( \beta^2 - 1 ) +
\Bigl[ \frac{1}{4}\,\lpar M^2 + \Gamma^2 \rpar - \mu\,\lpar
\mu + \frac{\Gamma\,\gamma}{M} \rpar\,\Bigr]\,\beta = 0.
\eq
For $\beta_{\rm min} < \beta < \beta_{\rm max}$ we have that $\chi^{(1,2)}$
never cross the positive imaginary axis. Furthermore, we compare
$\chi^{(1,2)}(0)$ with $\chi^{(1,2)}(\Gamma)$ at fixed $t \in [0,1]$ and
replace $\ln \to \ln^-$ when (always at fixed $t$) $\chi^{(1,2)}(\Gamma)$
crosses the negative real axis for some value $\Gamma_c$. In the example
that we are considering, illustrated in \fig{c_def}, everything is 
particularly simple since $\Imb\,\chi(\Reb\,\chi)$ is always a straight 
line but our recipe works, as well, in the general case and allows for a 
straightforward algorithmic implementation.

\begin{figure}[h]
\begin{tabular}{ccc}
\epsfig{file=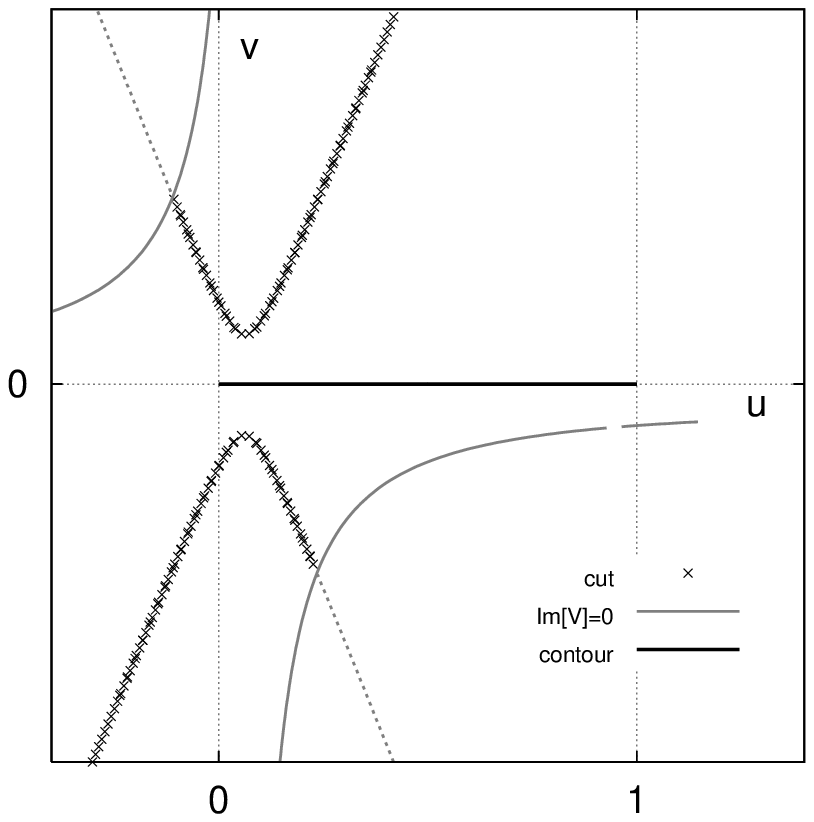, scale=0.6}
\hspace{-3cm}
&
\epsfig{file=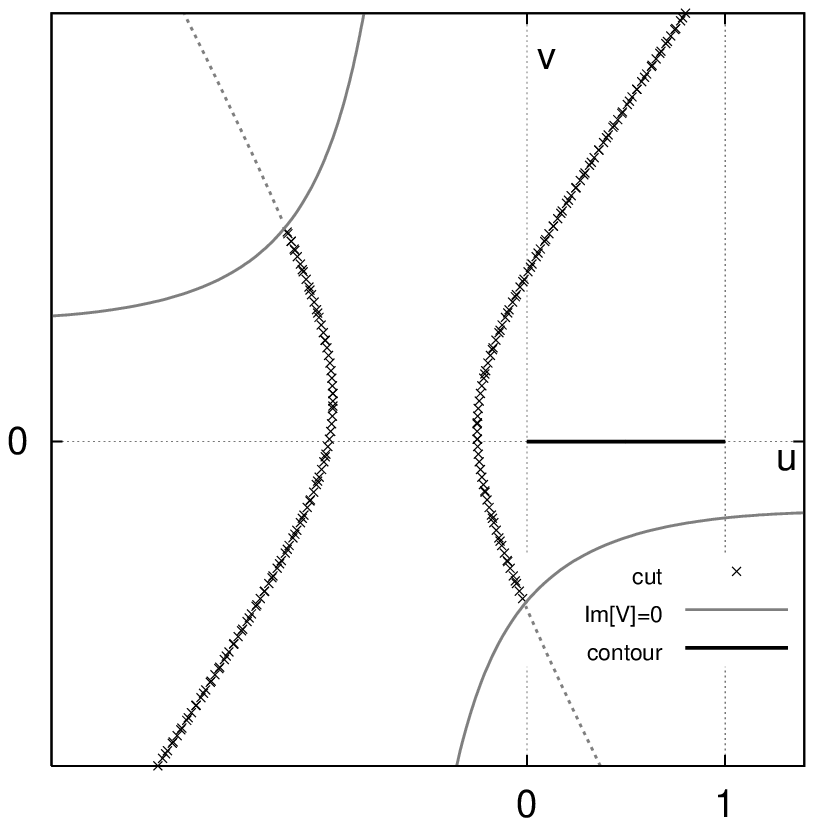, scale=0.6}
\hspace{-3cm}
&
\epsfig{file=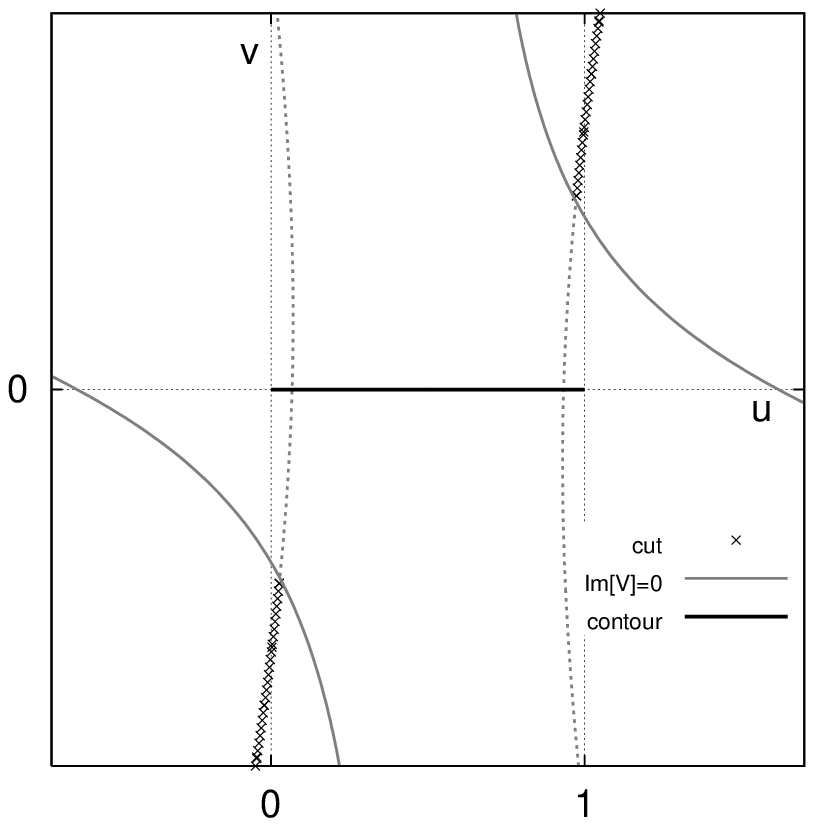, scale=0.6}
\\
\epsfig{file=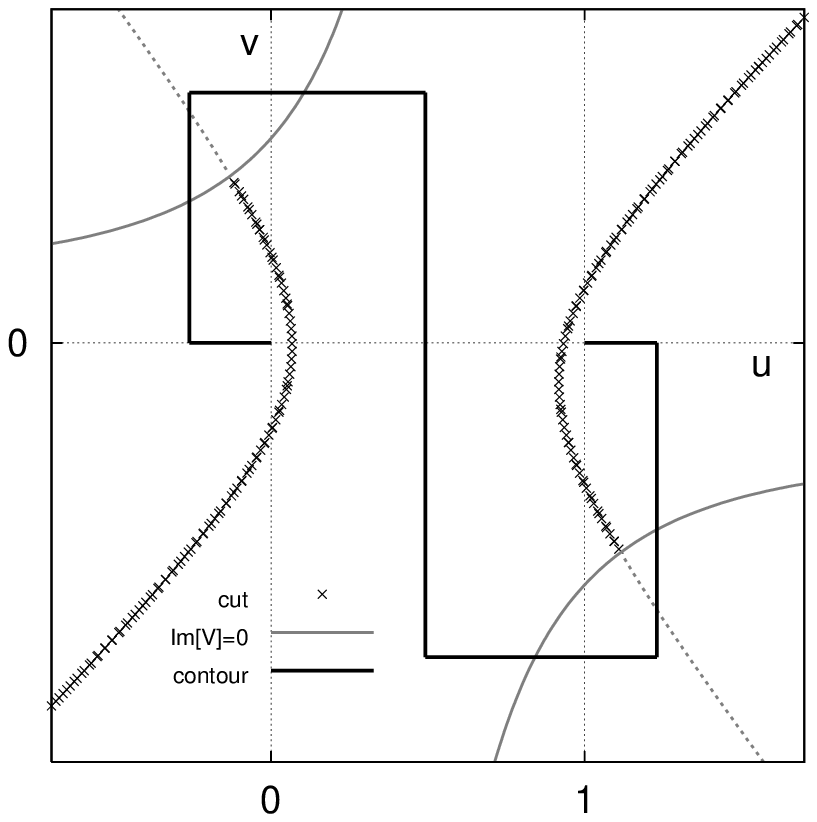, scale=0.6}
\hspace{-3cm}
&
\epsfig{file=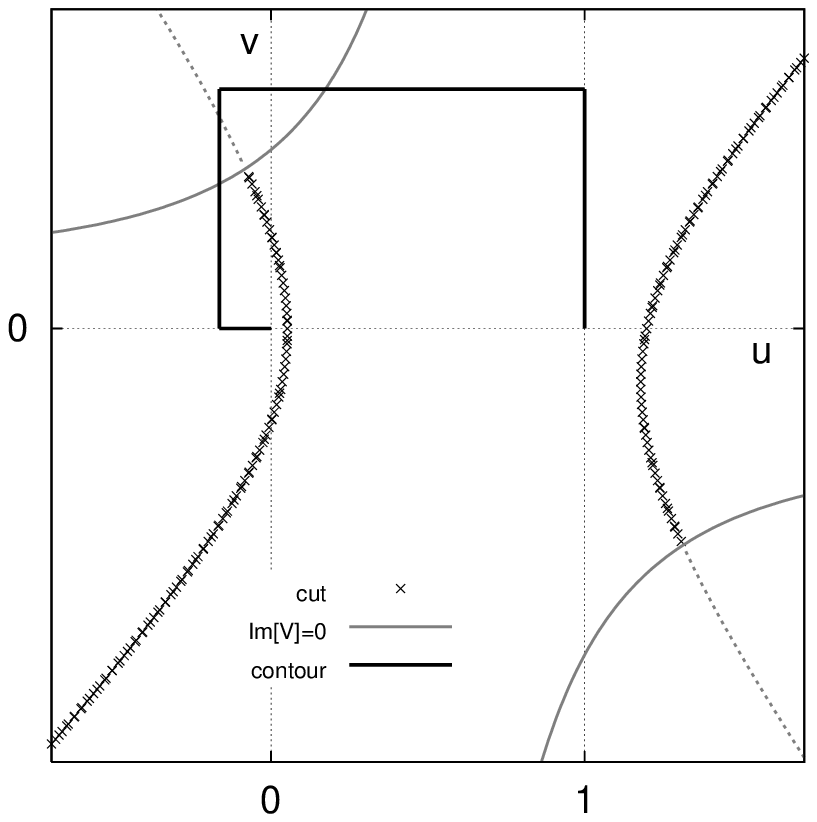, scale=0.6}
\hspace{-3cm}
&
\epsfig{file=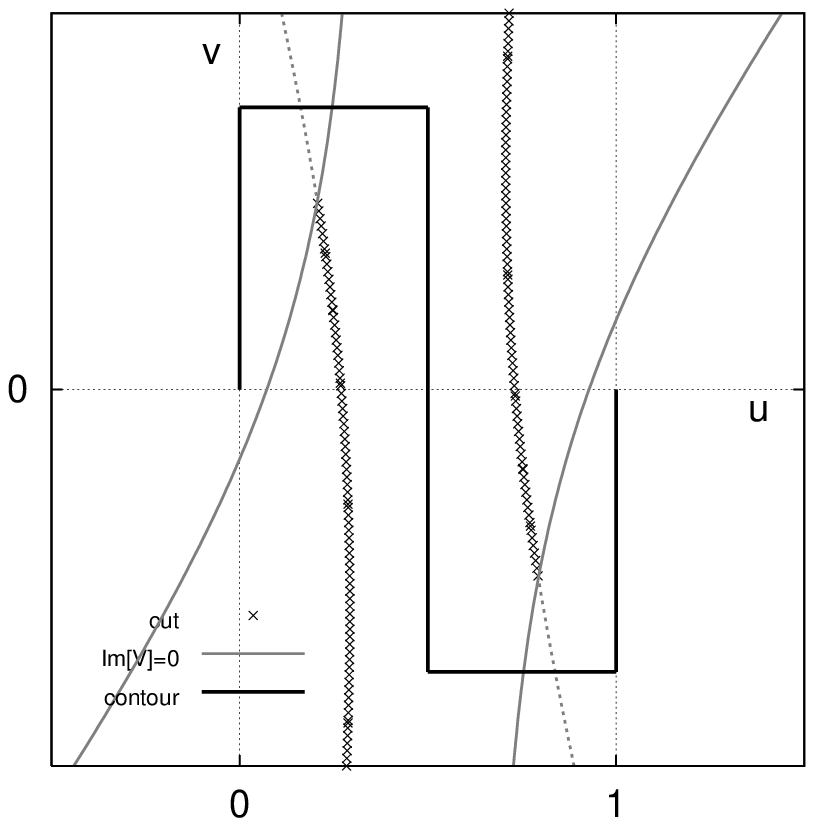, scale=0.6}
\end{tabular}
\vspace{-0.5cm}
\caption{Examples of deformation in the $x$-complex plane ($x=u+iv$) of the 
integration contour $[0,1]$ for integral of $\ln^-V=\ln^-(ax^2+bx+c)$.}
\vspace{-0.1cm}
\label{defor}
\end{figure}

For a general recipe of contour deformation, we proceed by analyzing the case 
where a Feynman diagram can be written as:
\bq
\int_0^1\!\!dx_1 \cdots dx_n\,
\sum_{i}\,A_i(x_1,\dots,x_n)\,\ln V_i(x_1,\dots,x_n),
\label{quad}
\eq
where $V_i$ are multivariate polynomials in $x_1,\dots,x_n$, at most quadratic 
in each variable (note that all one-loop diagrams can be written according to
\eqn{quad}, see Ref.~\cite{Ferroglia:2002mz}); actually the procedure works as 
well when each $V_i$ is a quadratic form in, at least, one variable.
For each term in the sum, we select one variable $x \equiv x_i$ (among 
$x_1,\dots,x_n$) and study the analytical continuation (assuming that 
${\rm Im}[V]_{\rm real\;masses} < 0$)
\bq
\ln V \to \ln^- V
\qquad
\mbox{ with }
\qquad
V= a\,x^2 + b\,x + c,
\eq
where $a,b,c$ are polynomials in the remaining Feynman variables.
The idea is to deform only the $x$ integration contour into the complex plane 
(when needed) while keeping all other variables 
($x_1,\dots,x_{i-1},x_{i+1},\dots,x_n$) on the real axis.
We define 
\bq
a=a_r+i\,a_i
\qquad
b=b_r+i\,b_i
\qquad
c=c_r+i\,c_i
\qquad\quad
x=u+i\,v.
\eq
The real and imaginary parts of $V$ are then given by:
\bqa
{\rm Re}V
&=& 
a_r\,u^2 - 2,a_i\,u\,v - a_r\,v^2 + b_r\,u - b_i\,v + c_r 
\nl
&=&
a_r\,(u-u_c)^2 - 2\,a_i\,(u-u_c)\,(v-v_c) - a_r\,(v-v_c)^2 + \delta_r,
\nl
{\rm Im}V
&=& 
a_i\,u^2 + 2,a_r\,u\,v - a_i\,v^2 + b_i\,u + b_r\,v + c_i
\nl
&=&
a_i\,(u-u_c)^2 + 2\,a_r\,(u-u_c)\,(v-v_c) - a_i\,(v-v_c)^2 + \delta_i,
\eqa
where we introduced the following auxiliary variables:
\bq
x_c= u_c + i\,v_c= -\,\frac{a^*\,b}{2\,|a|^2},
\qquad\qquad
\delta= \delta_r + i\,\delta_i= c - \frac{a^*\,b^2}{4\,|a|^2}.
\eq
The curves ${\rm Re}V=0$ and ${\rm Im}V=0$ are hyperbolas with center 
in $x_c$. We also define an auxiliary function, 
\bq
U= - a_i\,{\rm Re}V + a_r\,{\rm Im}V= 
2\,|a|^2\,(u-u_c)\,(v-v_c) + a_r\,\delta_i - a_i\,\delta_r.
\eq
The curve $U=0$ is again an hyperbola with center in $x_c$ and asymptotes
parallel to the $u$ and $v$ axes.
The branch-cut of $\ln^- V$ in the $x$ complex plane is defined by:
\bq
\mbox{cut}:
\qquad
{\rm Re}V=0 \;\,\&\,\; {\rm Im}V \geq 0 
\qquad
\Longleftrightarrow
\qquad
\left\{
\ba{c}
{\rm Re}V=0\;\,\&\,\;U\geq 0, \quad\mbox{ if } a_r>0, \\
{\rm Re}V=0\;\,\&\,\;U\leq 0, \quad\mbox{ if } a_r<0.
\ea
\right.
\eq
First we study the intersection of the curve ${\rm Re}V=0$ with the real $u\,$ 
axis. If $\Delta= b_r^2-4\,a_r\,c_r<0$, the hyperbola ${\rm Re}V= 0$ never crosses 
the real axis and there is no need of contour deformation (this is the first case 
of \fig{defor}).
If $\Delta \geq 0$, the intersections are given by:
\bq
{\rm Re}V=0\;\,\&\,\;v=0 
\qquad\Longrightarrow\qquad
u= u_0^\pm= \frac{-b_r\pm\sqrt{\Delta}}{2\,a_r}.
\eq
If both solutions are not in $[0,1]$ a distortion is not needed (second plot 
in \fig{defor}).
Even if $u_0^\pm \in [0,1]$, it can happen that the intersection occurs 
for ${\rm Im}V < 0$ (as in the third case of \fig{defor}) and there is no need 
of deformation to avoid the cut. 
In order to understand whether it occurs, we can study where the zeros 
of $V$ (${\rm Re}V= {\rm Im}V=0$) are lying.
This system of equations has always two and only two solutions 
$x_\pm=u_\pm+i\,v_\pm$, whose real and imaginary parts are given by:
\bq
u_\pm= u_c \pm \sqrt{ |\sigma| + \sigma_r },
\qquad\quad
v_\pm= v_c \pm\,\sign{\sigma_i}\sqrt{ |\sigma| - \sigma_r },
\qquad\qquad
\sigma= -\,\frac{a^*\,\delta}{2\,|a|^2}.
\eq
Note that these points are also solution of the equation $U= 0$ and (because of the 
simple form of $U$) we can conclude that:
\bq
\mbox{cut}\!:
\;\;
{\rm Re}V\!=0\;\,\&\,\;{\rm Im}V \geq 0 
\;\;
\Leftrightarrow
\,
\left\{
\ba{c}
{\rm Re}V\!=0\;\,\&\,\;U\geq 0
\;\;
\Leftrightarrow
\;\;
{\rm Re}V\!=0\;\,\&
\left\{
\ba{c}
v\geq v_+ \;\mbox{ if }\; u \!>\! u_c \\
v\leq v_- \;\mbox{ if }\; u \!<\! u_c \\
\ea
\right.
\;
\mbox{ if } a_r \!>\! 0,
\\
{\rm Re}V\!=0\;\,\&\,\;U\leq 0
\;\;
\Leftrightarrow
\;\;
{\rm Re}V\!=0\;\,\&
\left\{
\ba{c}
v\leq v_+ \;\mbox{ if }\; u \!>\! u_c \\
v\geq v_- \;\mbox{ if }\; u \!<\! u_c \\
\ea
\right.
\;
\mbox{ if } a_r \!<\! 0.
\ea
\right.
\eq
At this point we have all information to fix the new integration contour, 
starting from $x=0$ and ending at $x=1$ without crossing the branch-cut.
Of course, as long as the cut is not crossed, all integration contours are 
equivalent and give the same result: it may fairly be said that we have
some freedom in defining the deformation and that, at the same time, we can 
control the correctness of the result by using different paths.
The general situation is depicted in the fourth plot of \fig{defor} and 
the new integration contour is defined by seven segments:
\bqa
(1) &\quad& x \,=\, - \alpha_1\,t, \nl
(2) &\quad& x \,=\, - \alpha_1 + i\,\beta_1\,t, \nl
(3) &\quad& x \,=\, - \alpha_1\,(1-t) + \alpha_c\,t + i\,\beta_1, \nl
(4) &\quad& x \,=\,   \alpha_c + i\,\beta_1\,(1-t) + \beta_2\,t, \nl
(5) &\quad& x \,=\,   \alpha_c\,(1-t) + (1+\alpha_2)\,t + i\,\beta_2, \nl
(6) &\quad& x \,=\,   1 + \alpha_2 + i\,\beta_2\,(1-t), \nl
(7) &\quad& x \,=\,   (1+\alpha_2)\,(1-t) + t.
\label{newpath}
\eqa
The coefficients $\alpha_1$, $\beta_1$, $\alpha_2$, $\beta_1$ and $\alpha_c$ 
can be fixed according to the principle of the minimal deformation 
to avoid crossing the cut. This gives the following conditions:
\bqa
&&
\ba{lccccl}
\;\;\; \alpha_1 > \;\;\; |u_-| 
&
\quad {\rm if} \quad 
&
0 \leq u_0^- \leq 1
&
\quad\,\&\,\quad
&
 u_- \leq 0, 
&
\qquad \alpha_1= 0 \quad {\rm otherwise},
\\[+0.2cm]
\left\{
\ba{l}
\beta_1 > \;\;\; |v_-| \\
\beta_1 < - \, |v_-|   
\ea
\right.
&
\quad{\rm if}\quad
&
0 \leq u_0^- \leq 1 
&
\quad\,\&\,\quad
&
\ba{l}
v_- \geq 0 \;\,\&\,\; a_r>0, \\
v_- \leq 0 \;\,\&\,\; a_r<0,
\ea
&
\qquad \beta_1= 0 \quad {\rm otherwise},
\\[+0.4cm]
\;\;\; \alpha_2 > \;\;\; |u_+| 
&
\quad {\rm if} \quad
&
0 \leq u_0^+ \leq 1
&
\quad\,\&\,\quad
&
u_+ \geq 1, 
&
\qquad \alpha_2= 0 \quad {\rm otherwise},
\\[+0.2cm]
\left\{
\ba{ll}
\beta_2 < - \,   |v_+| \\
\beta_2 > \;\;\; |v_+| 
\ea
\right.
&
\quad{\rm if}\quad
&
0 \leq u_0^+ \leq 1
& 
\quad \,\&\,\quad
&
\ba{ll}
v_+ \leq 0 \;\,\&\,\; a_r>0, \\
v_+ \geq 0 \;\,\&\,\; a_r<0,
\ea
&
\qquad \beta_2= 0 \quad {\rm otherwise},
\\[+0.4cm]
\left\{
\ba{ll}
\alpha_c = \;\;\;\;   u_c \\
\alpha_c = \;\;\;\;\; 0   \\
\alpha_c = \;\;\;\;\; 1   \\
\ea
\right.
&
\quad{\rm if}\quad
&
\ba{l}
0     \leq u_0^- \leq u_0^+ \leq 1,    \\
u_0^- \leq  0    \leq u_0^+ \leq 1,    \\
0     \leq u_0^- \leq 1     \leq u_0^+,
\ea
&&&
\qquad \alpha_c= 0 \quad {\rm otherwise}.
\ea
\label{coeff}
\eqa
The case where all coefficients vanish corresponds to non-deformation:
in this case the paths in \eqn{newpath} are 
\bq
(1),\,(2),\,(3),\,(4)\;\; x=0,\quad(5)\;\;x=t,\quad(6),\,(7)\;\;x=1,
\eq
and refer to $ 0 \leq x \leq 1$ on the real axis.
Note that the case $\Delta < 0$ (where $u_0^\pm$ are not defined) belongs to 
this class.
This general recipe for contour deformation works also in special cases 
where not all segments are needed.
For example, in the fifth plot of \fig{defor} the new contour consists of 
only four segments: using in \eqn{newpath} the conditions of \eqn{coeff} 
($\alpha_2=\beta_2=0,\alpha_c=1$), the last three segments reduce in this 
case to the point $x=1$.

We can now consider in this framework the example of a $B_0$ function with two 
equal masses. In this case we have:
\bq
V(x) = 
-s_\ssP\,x\,(1-x)+m^2 = 
(-M^2+i\,M\,\Gamma)\,x\,(1-x)+\mu^2-i\,\mu\,\gamma,
\eq
which corresponds to 
\bqa
&&
x_c= \frac{1}{2}, \qquad\quad
\delta= 
c-\frac{a}{4}= 
\mu^2 - \frac{M^2}{2} - i\,\Big[ \mu\,\gamma + \frac{M\,\Gamma}{2} \Big], 
\nl
&&
\sigma = 
\frac{1}{8} -  \frac{a^*c}{2|a|^2} = 
\frac{1}{8} - \frac{\mu}{2M}\frac{\mu M+\gamma\Gamma}{M^2+\Gamma^2}
+ i\frac{\mu}{2M}\frac{\mu\Gamma -M\gamma}{M^2+\Gamma^2}.
\eqa
Since $a_r > 0$, the cut crosses the segment $[0,1]$ when $\sigma_i\leq 0$ 
(implying that $v_-\geq 0$, $v_+\leq 0$), a situation which occurs for 
$\mu\Gamma-M\gamma\geq 0$.
It can be verified by explicit calculation that, in this case, we always have 
$0\leq u_\pm\leq 1$, which corresponds to the situation depicted in the last 
diagram of \fig{defor} and the deformation requires five segments 
($\alpha_1=\alpha_2=0$, i.e. the first and the last segment in \eqn{newpath} 
reduce to a point).
\subsection{Differential operators in a complex domain \label{BSTc}}
The procedure of analytical continuation at the basis of \eqn{quad} deserves
an additional comment: how can we apply a differential operator at the
integrand level, in order to get \eqn{quad}? As an example we consider 
the integral of \eqn{eqdefC0} on which we want to apply the BST algorithm:
\bq
C_0 = \intfxy{x}{y}\,\chi^{-1+\ep/2}(x,y), \qquad \ep \to 0^+,
\eq
where $\chi$ is a quadratic form. As we know \eqn{quad} follows from BST
functional relation; in order to apply the BST algorithm in the 
complex domain ($\chi \in C[x,y]$) we introduce
\bq
\Bigl[ \chi \Bigr]_{\pm}^{\mu} = \exp \lpar \mu\,\ln^{\pm}\,\chi\rpar,
\eq
and distort the integration path so that it never crosses the positive
imaginary axis of $\chi^{-}$ (or the negative imaginary axis of $\chi^{+}$ ). 
The ($-$) analytical continuation of $C_0$ is defined by
\bq
C_0^- = \int_{\Lambda = 0}\,dx dy\,\Bigl[ \chi(x,y) \Bigr]_-^{-1+\ep/2},
\eq
where $\Lambda(x,y) = 0$ is the implicit equation for the integration contour.
In practice we change variables, $x= \alpha_i\,t+\beta_i$ 
($t \in [0,1]$) with $i=1,\dots,n$, n being the number of segments needed to 
avoid crossing the cut (e.g. see \eqn{newpath}).
The BST functional relation~\cite{Ferroglia:2002mz} for quadratic forms 
and the corresponding linear differential operator are
\bq
\chi^{\mu}\lpar [x] \rpar = 
{\cal D}\lpar \mu\,,\,[x]\,,\,[\partial_x]\rpar\,
\chi^{\mu+1}\lpar [x] \rpar, \qquad
{\cal D} = \frac{1}{B}\,\Bigl[ 1 - \frac{1}{2\,(\mu+1)}\,\sum_{i=1}^n\,
\lpar x_i - X_i\rpar\,\partial_{x_i} \Bigr],
\label{BSTexp}
\eq
where $[x] = x_1\,,\dots\,,x_n$. 
Consider the following integral,
\bq
F_- = \int_{z_i\;\Gamma}^{z_f}\,dz\,\Bigl[\chi(z)\Bigr]_-^{\mu}\;,
\label{Fminus}
\eq
where $\Gamma$ is a curve connecting $z_i$ and $z_f$ which never
crosses the positive imaginary axis of $\chi$ for $z \in \Gamma$. 
Let $z_f$ be in the second quadrant and $z_i$ outside of it;
let $z_0$ be the point where $\Imb\,\chi(z_0) = 0$, $\Reb\,\chi(z_0) < 0$. 

Thanks to the $-\,$ prescription the integrand in \eqn{Fminus} is a continuous 
function of $z$ (for $z \in \Gamma$) and we can write
\bq
F_- = F^{(1)}_- + F^{(2)}_- = 
      \int_{z_i\;\Gamma}^{z_0}\,dz\,\Bigl[\chi(z)\Bigr]_-^{\mu} +
      \int_{z_0\;\Gamma}^{z_f}\,dz\,\Bigl[\chi(z)\Bigr]_-^{\mu}.
\eq
In the first integral $\bigl[\chi\bigr]_-^{\mu} = \chi^{\mu}$ and we can apply 
the BST relation of \eqn{BSTexp}. In the second one we find
\bqa
F^{(2)}_- &=& \int_{z_0\;\Gamma}^{z_f}\,dz\,\Bigl[\chi(z)\Bigr]_-^{\mu} 
=
 \int_{z_0\;\Gamma}^{z_f}\,dz\, \exp\{-\,2\,i\,\pi\,(\mu+1)\}\,\chi^{\mu}(z) 
\nl
{}&=&
 \int_{z_0\;\Gamma}^{z_f}\,dz\, \exp\{-\,2\,i\,\pi\,(\mu+1)\}\,
{\cal D}\lpar \mu,z,\partial_z\rpar\,\chi^{\mu+1}(z) 
=
 \int_{z_0\;\Gamma}^{z_f}\,dz\, {\cal D}\lpar \mu,z,\partial_z\rpar\,
\Bigl[ \chi(z)\Bigr]_-^{\mu+1},
\label{BSTcmplx}
\eqa
where we have used $\exp\{-2 i \pi \mu\} = \exp\{- 2 i \pi (\mu+1)\}$, showing
extension of the BST algorithm into the second sheet. Note that the BST 
relation in the second equality of \eqn{BSTcmplx} is of pure algebraic nature
and that the integration over $\Gamma$ will always be parametrized in terms
of a $\Gamma(t)\,:\,t \in R$ so that $\chi \in C[t]$ and ${\cal D} \in
C[t]\,< \partial_t >$.

Each segment of \eqn{newpath} is of the type considered in $F^{(1)}_-$, 
$F^{(2)}_-$ or $F_-$ and therefore we can apply the BST algorithm to $C_0^-$.
It goes without saying that this is the correct procedure, instead of applying 
BST first and performing analytical continuation only in a second step. 
The result reads as follows:
\bq
C_0^-= \sum_{i=1}^n\,\alpha_i\,\frac{J_i}{B_3},
\qquad
J_i = 
  \intsx{x}\!\int_0^{\alpha_i\,x+\beta_i}\!\!\!\!\!\!dy\,
  \ln\!\!^-\!\chi (\alpha_i x\!+\!\beta_i,y) 
- \frac{1}{2}\,\sum_{j=1}^{4}\!\int_0^{a_{ij}}\!\!\!dx\,A_{ij}\,\ln^- \chi_{ij} 
+ \frac{\alpha_i}{2} + \beta_i,
\eq
\bq
\chi_{i1} = \chi( \alpha_i + \beta_i, x ), \qquad 
\chi_{i2} = \chi( \beta_i, x ), \qquad 
\chi_{i3} = \chi( \alpha_ix + \beta_i, \alpha_ix + \beta_i ), \qquad 
\chi_{i4} = \chi( \alpha_ix + \beta_i, 0 ),
\eq
\bq
A_{i1} = \frac{\alpha_i\!+\!\beta_i\!-\!X}{\alpha_i},\quad
A_{i2} = -\,\frac{\beta_i\!-\!X}{\alpha_i},\quad
A_{i3} = X\!-\!Y,\quad
A_{i4} = Y,\quad
a_{i1}\!= \alpha_i\!+\!\beta_i, \quad
a_{i2}\!= \beta_i, \quad
a_{i3}\!= a_{i4}\!= 1,
\eq
where $B_3$ is the BST factor and  $X,Y$ are the BST co-factors.

With this example we have shown how to apply differential operators when
complex momenta are present: first the analytical continuation has to be
performed together with the deformation of the integration contour and just at
the end the differential operator can be correctly applied.

In conclusion we have shown a practical implementation of the concept that
the pole at the mass of a stable particle can move into other Riemann sheets 
where it describes an unstable particle.
It is worth noting that all cases encountered so far, where both the internal 
masses and the Mandelstam invariants are complex, have never been discussed in 
the literature, although this step represents the logical extension of the 
complex-mass scheme, allowing for a meaningful introduction of pseudo-observables.
\section{Including QED(QCD) corrections \label{theQs}}
In this section we will consider the inclusion of QED(QCD) corrections, both
virtual and real. The choice $s = \cph$ in \eqn{PO} is dictated by the request 
of a gauge independent definition of pseudo-observables and follows, once 
again, from Nielsen identities.
Consider a final state where the inclusion of real QED(QCD) corrections is
mandatory in order to obtain an infrared finite quantity, e. g. $i \to H \to 
\barb b$.
Here, at one-loop, we have wave-function renormalization factors for the 
external fermions and vertex corrections; the QED part generates, in the 
so-called $\lpar \ep\,,\,m_b \rpar$ regulator scheme (dimensional 
regularization for the infrared and masses for the collinear limit), a 
simple infrared pole and double as well as simple collinear logarithm. 
According to our recipe the QED(QCD) vertex correction should be evaluated 
at complex Higgs momentum squared.
Let us define
\bq
\cph = x_{\ssH}\,\rph, \quad
\beta^2_c = 1 - 4\,\frac{m^2_b}{\cph}, \quad
\beta^2 = 1 - 4\,\frac{m^2_b}{\rph}.
\eq
The residue of the infrared (virtual) pole reads as follows
\bq
R_{\rm virt} = \Reb\,\frac{\beta}{\beta_c}\,\Bigl[
\frac{\beta^2}{x_{\ssH}} + 2 - \frac{1}{x_{\ssH}}\Bigr]\,
\ln^{-}\frac{\beta_c - 1}{\beta_c + 1}.
\eq 
The infrared pole from real emission originates from the end-point 
singularity in the phase space integration, where $P_{\ssH} = p_b + p_{\barb}$
and $P^2_{\ssH}$ is arbitrary but real, unless one is willing to extend the
phase space definition into the complex plane where $\delta$ functions are
defined in terms of contour integrals~\cite{Durand}.
Using the most obvious choice, namely $P^2_{\ssH} = - \rph$ we obtain for the 
(real) infrared residue
\bq
R_{\rm real} = \lpar \beta^2 + 1 \rpar\,\ln\frac{1-\beta}{1+\beta}.
\eq
Therefore, as expected, cancellation of infrared divergences is spoiled 
by the need of defining virtual corrections at a complex value of $s$.
However, the fact that $Z^{-1/2}_{\ssH}(s)\,V_f(s)$ is gauge parameter 
independent only at the complex pole does not exclude a gauge independent 
sub-set of corrections that can be evaluated at arbitrary $s$. 
Consider the situation at the one-loop level; here, in front of the 
$\ord{\alpha}$ QED corrections we use $Z_{\ssH} = 1$ and the one-loop vertex, 
$V^{\QED}_{\barb b}$, is gauge independent for all values of $s$.
 If we introduce
\bq
Z_{\ssH} = 1 + \frac{g^2}{16\,\pi^2}\,\delta Z_{\ssH},
\eq
for the wave-function renormalization factor, we can use
\bq
S\lpar H_c \to \barb b\rpar = 
-\,\frac{g}{2}\,\frac{m_b}{\mw}\,\Bigl\{
1 + \frac{g^2}{16\,\pi^2}\,\Bigl[ V^{\EW}_{\barb b}(s_{\ssH}) -
\frac{1}{2}\,\delta Z_{\ssH}(\cph) \Bigr] + 
\frac{\alpha}{4\,\pi}\,V^{\QED}_{\barb b}(\rph) \Bigr\},
\eq
thus preserving gauge invariance without spoiling infrared safety.
Following a well established convention it is also convenient to define a 
{\em deconvoluted} pseudo-observable where QED(QCD) corrections are 
subtracted according to theory.

There is an intriguing alternative for the treatment of QED(QCD) corrections;
since the definition of the Higgs boson mass is not unique, we could keep 
it as a free parameter, $\mh$. 
Then, cancellation of infrared poles at the one-loop level requires
\bq
\cph = x_{\ssH}\,\mhs, \quad
\beta^2_c = 1 - 4\,\frac{m^2_b}{\cph}, \quad
\beta^2 = 1 - 4\,\frac{m^2_b}{\mhs},
\eq
and $R_{\rm virt} = R_{\rm real}$. Therefore, there is a value of $\mh$ which is
infrared safe,
\bq
\mhs = 
| \cph |\,\Bigl\{ 1  + 2\,\frac{m^2_b}{|\cph|}\,
\Bigl[ 1 - \srph\,\lpar \rph + \gamma^2_{\ssH}\rpar^{-1/2}\Bigr] +
\ord{m^4_b} \Bigr\}.
\eq

\begin{figure}[h]
\vspace{0.3cm}
$$
\begin{picture}(140,30)(0,0)
 \SetScale{0.8}
 \SetWidth{1.8}
 \DashLine(0,0)(40,0){3}        
 \ArrowLine(100,-35)(140,-35)   
 \ArrowLine(140,35)(100,35)     
 \Line(70,-17.5)(40,0)          
 \Line(70,17.5)(40,0)           
 \ArrowLine(70,17.5)(70,-17.5)  
 \ArrowLine(70,-17.5)(100,-35)  
 \Photon(100,-35)(100,35){2}{7} 
 \ArrowLine(100,35)(70,17.5)    
\end{picture}
\qquad\qquad
\begin{picture}(140,30)(0,0)
 \SetScale{0.8}
 \SetWidth{1.8}
 \DashLine(0,0)(40,0){3}        
 \ArrowLine(100,-35)(140,-35)   
 \ArrowLine(140,35)(100,35)     
 \Line(70,-17.5)(40,0)          
 \Line(70,17.5)(40,0)           
 \ArrowLine(70,17.5)(100,0)  
 \Line(70,-17.5)(100,-35)  
 \Photon(100,35)(100,0){2}{4} 
 \ArrowLine(100,0)(100,-35)
 \ArrowLine(100,35)(70,17.5)    
\end{picture}
$$
\vspace{0.5cm}
\caption[]{
Examples of mixed electroweak-QED two-loop diagrams contributing to $H \to \barb b$. 
The solid lines, attached to the Higgs boson (dash-line) represent
$Z/\phi^0$ or $W/\phi$ fields.}
\label{MTL}
\end{figure}
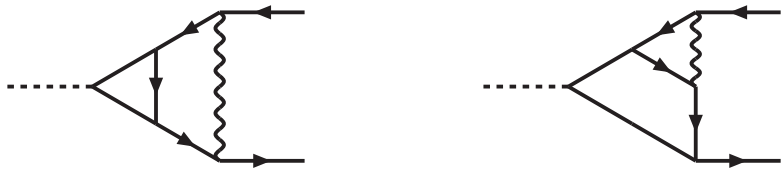
The main question in establishing the consistency of the procedure, 
definition of QED(QCD) corrections and their subsequent deconvolution, 
is about the extendability to higher orders. 
Using the following expansions
\bq
Z_{\ssH} = 1 + \sum_{n=1}^{\infty}\,\frac{g^{2 n}}{16\,\pi^2}\,
\delta Z^{(n)}_{\ssH},
\quad
V_{\barb b} = \sum_{n=1}^{\infty}\,
\frac{g^{2 n -1}}{16\,\pi^2}\,V^{(n-1)}_{\barb b},
\quad
V^{(0)}_{\barb b} = -\,\frac{m_b}{2\,\mw}
\eq
and working at $\ord{g^5}$ we will have terms like
\bq
\delta Z^{(1)}_{\ssH}\,V^{(1)\,;\,\QED}_{\barb b},
\label{mixedZ}
\eq
which are of the mixed type, electroweak-QED, and where $Z_{\ssH}$ 
cannot be evaluated at arbitrary values of $s$. 
In \eqn{mixedZ} $V^{(1)\,;\,\QED}_{\barb b}$ is the one-loop QED triangle 
contributing to $H \to \barb b$. 

\begin{figure}[h]
\vspace{0.2cm}
$$
\raisebox{0.1cm}{
\begin{picture}(110,30)(0,0)
 \SetScale{0.8}
 \SetWidth{1.8}
 \DashLine(0,0)(40,0){3}        
 \Line(140,-35)(100,-35)   
 \Line(140,35)(100,35)     
 \Line(70,-17.5)(40,0)          
 \Line(70,17.5)(40,0)           
 \Line(70,17.5)(70,-17.5)  
 \Line(70,-17.5)(100,-35)  
 \Photon(100,-35)(100,35){2}{7} 
 \Line(100,35)(70,17.5)    
\end{picture}}
= 
\quad
\raisebox{0.1cm}{
\begin{picture}(80,30)(0,0)
 \SetScale{0.6}
 \SetWidth{2}
 \DashLine(0,0)(40,0){3}        
 \Line(120,-35)(80,-35)   
 \Line(120,35)(80,35)     
 \Line(80,-35)(40,0)          
 \Line(80,35)(40,0)           
 \Line(80,-35)(80,35)
\end{picture}}
\otimes
\;
\raisebox{0.1cm}{
\begin{picture}(80,30)(0,0)
 \SetScale{0.6}
 \SetWidth{2}
 \DashLine(0,0)(40,0){3}        
 \Line(120,-35)(80,-35)   
 \Line(120,35)(80,35)     
 \Line(80,-35)(40,0)          
 \Line(80,35)(40,0)           
 \Photon(80,-35)(80,35){2}{7}
\end{picture}}
\quad
+
\quad
\mbox{IR finite}
$$
\vspace{0.2cm}
\caption[]{Infrared decomposition of a mixed electroweak-QED two-loop diagram. Here the
scalar case is presented, i.e. the spin structure has been completely neglected
and, for instance, the wavy line represents a scalar massless line.}
\label{IRdec}
\end{figure}
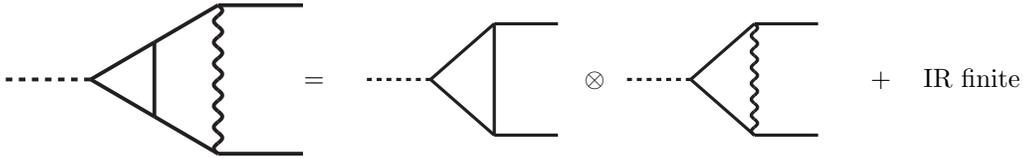

However, we also have mixed two-loop diagrams, as given in \fig{MTL}.
There is a well-known identity which allows us to extract the infrared 
behavior of these two-loop diagram, in terms of the product of two 
one-loop vertices plus an infrared finite reminder (see \cite{Passarino:2006gv} 
for the explicit decomposition in the scalar case). 
The decomposition for the scalar case is illustrated in \fig{IRdec} where 
the external lines in both one-loop vertices are on-shell. 
By scalar we mean those contributions that do not have powers of the 
integration momentum in the numerator. 

\begin{figure}[ht!]
\vspace{0.2cm}
$$
\raisebox{0.1cm}{
\begin{picture}(110,30)(0,0)
 \SetScale{0.8}
 \SetWidth{1.8}
 \DashLine(0,0)(40,0){3}        
 \ArrowLine(100,-35)(140,-35)   
 \ArrowLine(140,35)(100,35)     
 \Line(70,-17.5)(40,0)          
 \Line(70,17.5)(40,0)           
 \ArrowLine(70,17.5)(70,-17.5)  
 \ArrowLine(70,-17.5)(100,-35)  
 \Photon(100,-35)(100,35){2}{7} 
 \ArrowLine(100,35)(70,17.5)    
 \Text(110,35)[cb]{$v$}
 \Text(110,-40)[cb]{$\baru$}
\end{picture}}
\quad
=
\quad
\frac{g^2 \stw^2}{16\,\pi^2}\,Q^2_b\,( \mhs - 2\,m^2_b )
\;
\raisebox{0.1cm}{
\begin{picture}(80,30)(0,0)
 \SetScale{0.6}
 \SetWidth{2}
 \DashLine(0,0)(40,0){3}        
 \ArrowLine(80,-35)(120,-35)   
 \ArrowLine(120,35)(80,35)     
 \Line(80,-35)(40,0)          
 \Line(80,35)(40,0)           
 \ArrowLine(80,35)(80,-35)
 \Text(70,28)[cb]{$v$}
 \Text(70,-33)[cb]{$\baru$}
\end{picture}}
\!\!
\otimes
\;\;
C^{\rm IR}_0
\qquad + \quad \hbox{IR finite}
$$
\vspace{0.2cm}
\caption[]{Infrared decomposition of a mixed electroweak-QED two-loop amplitude. 
Solid lines represent $Z, \phi^0$ or $W, \phi$ particles; Dirac spinors for 
the external lines are included and $C^{\rm IR}_0$ is the scalar, 
infrared divergent, three-point function.}
\label{IRampdec}
\end{figure}
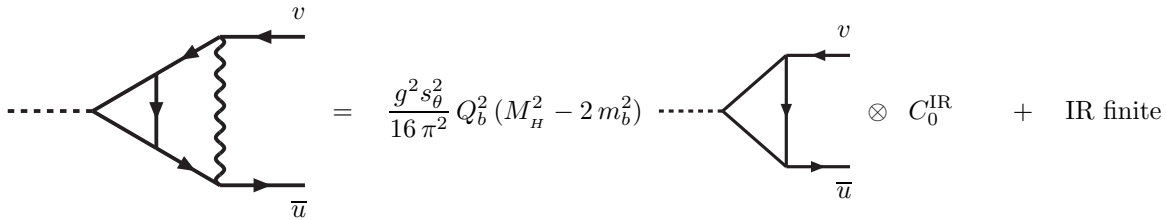

As we have seen, the identity holds at the amplitude level, reflecting 
the factorization of virtual infrared corrections and the fact that virtual 
infrared poles are always coming from $C_0\,$-functions (the scalar ones). 
These identities follow from the fact that any diagram with an infrared 
photon line of momentum $q_i + K$, where $K$ is a certain combination of 
external momenta as well as of the other loop momenta, gives an infrared 
divergence equivalent to the same diagram evaluated at $q_i = -K$. 
We thus see that the infrared decomposition into products of tensor 
integrals times infrared $C_0\,$-functions follows trivially.
The explicit form of infrared factorization, at the amplitude level, is 
illustrated in \fig{IRampdec} which shows a class of diagrams contributing to
the two-loop amplitude for $H \to \barb b$. 

When added to the contribution coming from 
$\delta Z^{(1)}_{\ssH}\,V^{(1)}_{\barb b}$, the combination
\bq
V^{(1) \EW}_{\barb b}(s) - \frac{1}{2}\,\delta Z^{(1)}_{\ssH}(s)
\label{combi}
\eq
arises naturally in front of an infrared $C_0$. 
Therefore, our recipe will be to evaluate \eqn{combi} at $s = \cph$ while 
keeping the remaining QED-like $C_0\,$functions (the infrared divergent 
ones) at $s= \rph$. 
The difference with the original diagram is non resonant and mixes with 
infrared divergent background contributions, e.g. from boxes.

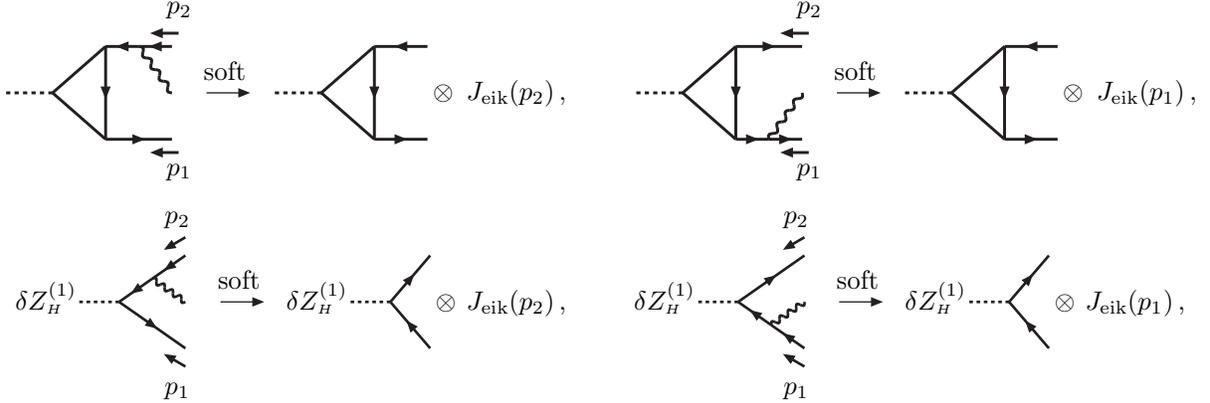
\begin{figure}[h]
\vspace{0.2cm}
\bqas
\raisebox{0.1cm}{
\begin{picture}(70,30)(5,0)
 \SetScale{0.5}
 \SetWidth{2.2}
 \DashLine(5,0)(40,0){3}        
 \ArrowLine(80,-35)(130,-35)   
 \ArrowLine(130,35)(105,35)     
 \ArrowLine(105,35)(80,35)
 \Photon(105,35)(130,0){2}{4}
 \Line(80,-35)(40,0)          
 \Line(80,35)(40,0)           
 \ArrowLine(80,35)(80,-35)
 \LongArrow(135,45)(115,45)       \Text(68,28)[cb]{$p_2$}
 \LongArrow(135,-45)(115,-45)     \Text(68,-33)[cb]{$p_1$}
\end{picture}}
\raisebox{0.1cm}{
\begin{picture}(15,0)(0,0)
 \LongArrow(0,0)(15,0)
 \Text(7,5)[cb]{soft}
\end{picture}}
\quad
\raisebox{0.1cm}{
\begin{picture}(55,30)(5,0)
 \SetScale{0.5}
 \SetWidth{2.2}
 \DashLine(5,0)(40,0){3}        
 \ArrowLine(80,-35)(120,-35)   
 \ArrowLine(120,35)(80,35)     
 \Line(80,-35)(40,0)          
 \Line(80,35)(40,0)           
 \ArrowLine(80,35)(80,-35)
\end{picture}}
\otimes\;
J_{\rm eik}(p_2)\,,
&\qquad&
\raisebox{0.1cm}{
\begin{picture}(70,30)(5,0)
 \SetScale{0.5}
 \SetWidth{2.2}
 \DashLine(5,0)(40,0){3}        
 \ArrowLine(80,35)(130,35)   
 \ArrowLine(105,-35)(130,-35)     
 \ArrowLine(80,-35)(105,-35)
 \Photon(105,-35)(130,0){2}{4}
 \Line(80,-35)(40,0)          
 \Line(80,35)(40,0)           
 \ArrowLine(80,35)(80,-35)
 \LongArrow(135,45)(115,45)       \Text(68,28)[cb]{$p_2$}
 \LongArrow(135,-45)(115,-45)     \Text(68,-33)[cb]{$p_1$}
\end{picture}}
\raisebox{0.1cm}{
\begin{picture}(15,0)(0,0)
 \LongArrow(0,0)(15,0)
 \Text(7,5)[cb]{soft}
\end{picture}}
\quad
\raisebox{0.1cm}{
\begin{picture}(55,30)(5,0)
 \SetScale{0.5}
 \SetWidth{2.2}
 \DashLine(5,0)(40,0){3}        
 \ArrowLine(80,-35)(120,-35)   
 \ArrowLine(120,35)(80,35)     
 \Line(80,-35)(40,0)          
 \Line(80,35)(40,0)           
 \ArrowLine(80,35)(80,-35)
\end{picture}}
\otimes\;
J_{\rm eik}(p_1)\,,
\\[1.4cm]
\delta Z^{(1)}_{\ssH}\;
\raisebox{0.1cm}{
\begin{picture}(45,30)(10,0)
 \SetScale{0.5}
 \SetWidth{2}
 \DashLine(10,0)(40,0){3}        
 \ArrowLine(40,0)(90,-35)   
 \ArrowLine(90,35)(65,17.5)     
 \ArrowLine(65,17.5)(40,0)
 \Photon(65,17.5)(90,0){2}{4}
 \LongArrow(90,48.75)(75,40)       \Text(42,28)[cb]{$p_2$}
 \LongArrow(90,-48.75)(75,-40)     \Text(42,-37)[cb]{$p_1$}
\end{picture}}
\raisebox{0.1cm}{
\begin{picture}(15,0)(0,0)
 \LongArrow(0,0)(15,0)
 \Text(7,5)[cb]{soft}
\end{picture}}
\quad
\delta Z^{(1)}_{\ssH}\;
\raisebox{0.1cm}{
\begin{picture}(25,30)(10,0)
 \SetScale{0.5}
 \SetWidth{2}
 \DashLine(10,0)(40,0){3}        
 \ArrowLine(40,0)(70,35)   
 \ArrowLine(70,-35)(40,0)
\end{picture}}
\otimes\,
J_{\rm eik}(p_2)\,,
&\qquad&
\delta Z^{(1)}_{\ssH}\;
\raisebox{0.1cm}{
\begin{picture}(45,30)(10,0)
 \SetScale{0.5}
 \SetWidth{2}
 \DashLine(10,0)(40,0){3}        
 \ArrowLine(40,0)(90,35)   
 \ArrowLine(90,-35)(65,-17.5)     
 \ArrowLine(65,-17.5)(40,0)
 \Photon(65,-17.5)(90,0){2}{4}
 \LongArrow(90,48.75)(75,40)       \Text(42,28)[cb]{$p_2$}
 \LongArrow(90,-48.75)(75,-40)     \Text(42,-37)[cb]{$p_1$}
\end{picture}}
\raisebox{0.1cm}{
\begin{picture}(15,0)(0,0)
 \LongArrow(0,0)(15,0)
 \Text(7,5)[cb]{soft}
\end{picture}}
\quad
\delta Z^{(1)}_{\ssH}\;
\raisebox{0.1cm}{
\begin{picture}(25,30)(10,0)
 \SetScale{0.5}
 \SetWidth{2}
 \DashLine(10,0)(40,0){3}        
 \ArrowLine(40,0)(70,35)   
 \ArrowLine(70,-35)(40,0)
\end{picture}}
\otimes\,
J_{\rm eik}(p_1)\,,
\eqas
\vspace{0.2cm}
\caption[]{
Examples of the infrared decomposition of $\ord{g^3}$ electroweak diagrams 
with real photon emission. The last term in the r.h.s of the equation is
the corresponding eikonal factor of \eqn{eikf}.
}
\label{RIRdec}
\end{figure}

For one-loop real emission we have diagrams as illustrated in 
the l.h.s. of \fig{RIRdec} where we use $p^2_{\ssH}= -\cph$ and where the 
infrared singularity arises from the end-point of phase space integration 
which is controlled by a real Higgs boson momentum. 
The corresponding amplitude is gauge independent by construction, a fact 
that can be easily seen in the infrared divergent soft approximation, 
the r.h.s. of \fig{RIRdec}, where we have introduced the eikonal 
factor 
\bq
J_{\rm eik}(p)= -\,Q_b\,\frac{\spro{p}{\ep}}{\spro{p}{k}}, 
\qquad \spro{\ep(k)}{k}= 0,
\label{eikf}
\eq
$\ep$ being the photon polarization. The vertex correction, first diagram in 
the r.h.s. of \fig{RIRdec}, when summed with 
$Z^{-1/2}_{\ssH}\,\otimes\,{\rm LO}$ gives a gauge invariant contribution if 
both are evaluated at the Higgs complex pole.  


There is another example where the introduction of QED corrections seems
to be controversial if only internal masses are made complex\footnote{This point
was raised by Thomas Binoth in one of our last conversations.}. Let us consider 
the pseudo-observable $\Gamma \lpar H \to W^+ W^-\rpar$, with on-shell 
external vector bosons and internal complex masses: the infrared behavior 
of the one-loop corrections is reducible to a scalar vertex
\bq
C_0\lpar -\mws\,,\,-\mws\,,\,-s\,,\,s_{\ssW}\,,0\,,s_{\ssW}\rpar,
\eq
where the difference $s_{\ssW} - \mws$ ($s_{\ssW}$ being the $W$ complex pole) acts
as an infrared regulator, removing the infrared virtual pole,
\bq
C_0\lpar -\mws\,,\,-\mws\,,\,-s\,,\,s_{\ssW}\,,0\,,s_{\ssW}\rpar =
\frac{2}{\beta_{\ssW}\,s}\,\ln\frac{\beta_{\ssW}+1}{\beta_{\ssW}-1}\,
\ln\frac{s_{\ssW}-\mws}{s} + \dots
\eq
where $\beta^2_{\ssW} = 1-4\,\mws/s$. If we continue the external masses the 
result is instead
\bq
C_0\lpar -s_{\ssW}\,,\,-s_{\ssW}\,,\,-s\,,\,s_{\ssW}\,,0\,,s_{\ssW}\rpar =
\frac{1}{\beta_{c\ssW}\,s}\,\ln^{-}\frac{\beta_{c\ssW}+1}{\beta_{c\ssW}-1}\,
\frac{1}{\bar\ep} + \hbox{IR finite},
\eq
where $\beta^2_{c\ssW} = 1-4\,s_{\ssW}/s$. Let us consider a realistic example, 
e.g. $gg \to 4\,$f of \fig{Multi}; for the complete process there is no 
problem at all because a photon attached to an internal $W$ boson line cannot 
be infrared divergent.
However, the goal is a breakdown of the full process into three components,
one of which is the pseudo-observable $\Gamma \lpar H \to W^+ W^-\rpar$;
in order to define $\Gamma\lpar H \to W^+ W^- (\gamma)\rpar$ it is 
important to control the cancellation between virtual and real infrared 
corrections and, in this case, the extension to external complex masses is 
more than an option. 
\section{Schemes \label{schemes}}
For processes which are relevant for the LHC and, in particular, for $H \to \barb b$, 
$\gamma\gamma$, $gg$ and $gg \to H$ etc, we define three different schemes
and compare their results. The schemes are:
\bei
\item the RMRP scheme which is the usual on-shell scheme where all
  masses and all Mandelstam invariants are real;
\item the CMRP scheme~\cite{Actis:2008uh}, the complex mass scheme with complex 
internal $W$ and $Z$ poles (extendable to top complex pole) but with real, 
external, on-shell Higgs, $W, Z\,$, etc. legs and with the standard LSZ wave-function 
renormalization;
\item the CMCP scheme, the (complete) complex mass scheme with complex,
external, Higgs ($W, Z$, etc.) where the LSZ procedure is carried out at the 
Higgs complex pole (on the second Riemann sheet).
\eei
The introduction of three different schemes does not reflect a theoretical
uncertainty; only the CMCP scheme is fully consistent and comparisons only
serve the purpose of quantifying deviations of more familiar schemes from 
the CMCP scheme.
\section{Numerical results \label{Nres}}
In this section we examine the numerical impact of computing Higgs
pseudo-observables at the Higgs complex pole (on the second Riemann sheet 
of the $S\,$-matrix).
We use the parametrization $\cph= \rph - i\,\srph\,\lgh$ for the Higgs complex pole
where now $\srph$ is an input parameter and $\lgh$ is computed
in the standard model. In this case $\srph$ plays the role of an
input parameter while we prefer $\brph$ of \eqn{MHdef} as conventional definition
of the Higgs boson {\em mass}. The results are compared with on-shell pseudo 
observables.

As input parameters for the numerical evaluation we have used the
following values
\[
\begin{array}{llll}
\mw = 80.398\,\GeV,  \;\; & \;\; 
\mz = 91.1876\,\GeV, \;\; & \;\;
\mt = 170.9\,\GeV,   \;\; & \;\;
\Gamma_{\ssW} = 2.093\,\GeV,\\ [0.3cm]
\gf = 1.16637\,\times\,10^{-5}\,\GeV^{-2},  \;\; & \;\; 
\alpha(0) = 1/137.0359911,                  \;\; & \;\; 
\alpha_{\ssS}\lpar \mz\rpar= 0.118,         \;\; & \;\; 
\Gamma_{\ssZ} = 2.4952\,\GeV.
\end{array}
\]
For the $W, Z$ complex poles we use
\bq
s_{\ssV}= \mu^2_{\ssV} - i\,\mu_{\ssV}\,\gamma_{\ssV},
\quad
\mu^2_{\ssV}= M^2_{\ssV} - \Gamma^2_{\ssV},
\quad
\mu_{\ssV}\gamma_{\ssV}=\,  
M_{\ssV}\,\Gamma_{\ssV}\,
\lpar 1 - \frac{1}{2}\,\frac{\Gamma^2_{\ssV}}{M^2_{\ssV}} \rpar.
\eq
In computing $H \to gg(gg \to H)$ we have used a running 
$\alpha_{\ssS}(\srph)\,$(CMRP) or $\alpha_{\ssS}(\sbrph)\,$(CMCP).
Results for the computed $\lgh$ are collected in \tabn{TgammaH}.
\begin{table}[h!]\centering
\setlength{\arraycolsep}{\tabcolsep}
\renewcommand\arraystretch{1.2}
\begin{tabular}{|c|c|c|c|c|c|c|c|c|}
\hline 
&&&&&&&& \\[-0.4cm]
$\srph\;$[GeV] &  $100$ & $120$ & $160$ & $170$ & $180$ & $200$ & $250$ & $400$  \\
&&&&&&&& \\[-0.4cm]
\hline
&&&&&&&& \\[-0.4cm]
$\lgh\,$[GeV] &
$0.051$ & $0.043$ & $0.105$ & $0.391$ & $0.637$ & $1.448$ & $4.296$ & $39.729$  \\
&&&&&&&& \\[-0.4cm]
\hline
&&&&&&&& \\[-0.4cm]
$\lgh\,$[GeV] &
$0.051$ & $0.043$ & $0.105$ & $0.391$ & $0.637$ & $1.448$ & $4.373$ & $39.829$  \\
&&&&&&&& \\[-0.4cm]
\hline
&&&&&&&& \\[-0.4cm]
$\lgh\,$[GeV] &
$0.051$ & $0.043$ & $0.105$ & $0.391$ & $0.637$ & $1.498$ & $5.069$ & $40.847$  
\\[-0.4cm]
&&&&&&&& \\
\hline
\end{tabular}
\caption[] {Standard model prediction for $\lgh$ in GeV as a function of
$\srph$. The first entry corresponds to a real on-shell top quark mass, 
the second entry to a top quark complex pole derived from 
$\Gamma_t =\Gamma^{\NLO}_t = 1.31\,$GeV and the last entry to a top quark complex 
pole derived from $\Gamma_t$ equal to the experimental upper bound of 
$13.1\,$GeV.}
\label{TgammaH}
\end{table}

For the evaluation of all one-loop functions in the CMCP scheme where, sometimes, 
a continuation to the second Riemann sheet is required we used both the analytical
results and the exact numerical integration; the two in-house (independent) libraries
return results in excellent agreement (typically on the sixth digit on one-loop
percentage radiative corrections). 
\subsection{Numerical differences between the CMRP and CMCP schemes \label{NDiff}}
For a better understanding of comparisons we define weak corrections to
$H \to \barb b$ as
\bq
\Delta_{\rm weak} = \sqrt{2}\,\frac{G_{\ssF}\,\brph}{\pi^2}\,
\lpar C_{\rm part} + B_{\rm part} + R \rpar,
\qquad
\brph = \mu_{\ssH}\,\lpar \rph + \gamma^2_{\ssH} \rpar^{1/2},
\label{parts}
\eq
separating the corrections into a part coming from three-point functions($C_{\rm part}$), 
two-point functions($B_{\rm part}$) and a rational term($R$). 
It is worth noting that there are, in general, strong cancellations among 
the three contributions: for instance, at $120\,$GeV we have a $C_{\rm part}$ 
of $-8.233\,\%$ from CMCP exact while the bracket in \eqn{parts} is $-0.790\,\%$.

Differences between the two schemes are roughly of $\ord{\lgh/\srph}$, 
as expected, and become significant above the $\bart t\,$-threshold 
where the Higgs boson width becomes larger and larger.
Since the width of an heavy Higgs boson is large it is natural to 
investigate the goodness of the separation of the production stage 
from the decay process. 
In general these stages are not independent and may be interconnected 
by radiative effects.
Our results confirm the theorem of Ref.~\cite{Fadin:1993kt}: 
radiative effects are not enhanced in totally inclusive pseudo-observables with 
respect to the naive $\ord{\lgh/\srph}$ argument, unless the Higgs boson is 
very heavy, in which case this ratio is large (at $\srph = 500\,$GeV it reaches 
$29\%$) and typical cancellations in the total weak correction factor are 
disturbed, increasing the effect.

To further understand the differences between the two schemes for high
values of $\srph$, we recall the well-known fact that the Higgs 
wave-function renormalization shows an inverse $\beta\,$-behavior at the
$W, Z$ threshold. In the two schemes, exactly at threshold, we will have
\bq
\beta^2= 1 - 4\,\frac{\mu^2_{\ssB} - i\,\gamma_{\ssB}\,\mu_{\ssB}}{m^2_{\ssH}}
\bmid_{\rm thr} = i\,\frac{\gamma_{\ssB}}{\mu_{\ssB}},
\qquad\quad
\beta^2_c= 1 - 4\,\frac{\mu^2_{\ssB} - i\,\gamma_{\ssB}\,\mu_{\ssB}}
{\rph - i\,\lgh\,\srph}
\bmid_{\rm thr} \sim i\,\frac{\mid 2\,\gamma_{\ssB} - \lgh\mid}
{2\,\mu_{\ssB}},
\eq
where $B= W, Z$. The parameter that regularizes the divergence is therefore
$\gamma_{\ssB} - \lgh/2$ with some sizable effect around the $ZZ$ 
threshold. To analyze differences between the CMRP and CMCP schemes we fix $\srph$
and compute $\lgh$; then we use \eqn{parts} and compare results
with the limit $\lgh = 0$.
Results are given in \tabn{CMRPvsCMCP} where we see variations induced 
by a finite $\lgh$.
\begin{table}[h!]\centering
\setlength{\arraycolsep}{\tabcolsep}
\renewcommand\arraystretch{1.2}
\begin{tabular}{|c|c|c|c|c|}
\hline 
&&&& \\[-0.4cm]
$\lgh/\srph$ & $C_{\rm part}$ & $B_{\rm part}$ & $R$ & tot \\
&&&& \\[-0.4cm]
\hline
&&&& \\[-0.4cm]
$0$        &  $-3.673$  &  $-1.999$  &  $+4.514$ & $-1.658$ \\
$0.03 $    &  $-3.760$  &  $-1.990$  &  $+4.009$ & $-1.741$ \\
&&&& \\[-0.4cm]
\hline
&&&& \\[-0.4cm]
$0$        &  $-0.308$  &  $-0.130$  &  $+3.450$ & $+3.058$ \\
$0.29$     &  $-0.986$  &  $+0.974$  &  $+2.714$ & $+2.702$ \\[-0.4cm]
&&&& \\
\hline
\end{tabular}
\caption[] {Variations (in percent) at $\srph = 300, 500\,$GeV in the components
of the total weak corrections to $H\to\bar{b}b$ according to \eqn{parts}.}
\label{CMRPvsCMCP}
\end{table}
\subsection{Testing the NWA approximation \label{TNWA}}
In order to analyze the quality of the NWA approximation (\sect{NWA}) we have 
considered the pure one-loop weak corrections to the decay width $H \to \barb b$. 
It turns out, that up to $\srph = 250\,$GeV (where $\lgh/\srph = 0.011$) the 
approximation is very good, less than $1\%$ (of a $\approx\,1\%$ correction). 
Note that analytical continuation of three-point functions in the
exact CMCP scheme is required above $220\,$GeV. 
\subsection{Complete set of results \label {CSR}}
As far as the Higgs boson production cross section in gluon-gluon fusion 
is concerned we find that the effect of replacing the on-shell scheme (for the 
external Higgs boson) with the complex-pole one is completely negligible (around $3-4$ 
per mill) for low values of the Higgs mass, a fact that is largely expected. Only 
for higher values, say starting from the $\bart t\,$threshold, where $\lgh$ becomes 
larger and larger, we reach sizable differences, above $10\%$ and rapidly 
increasing. 

\begin{itemize}
\item $\mathbf{H \to \gamma\gamma, gg, \;\; gg \to H}$
\end{itemize} 

\begin{figure}[ht!]
\begin{center}
\includegraphics[bb=0 0 567 384,width=12cm]{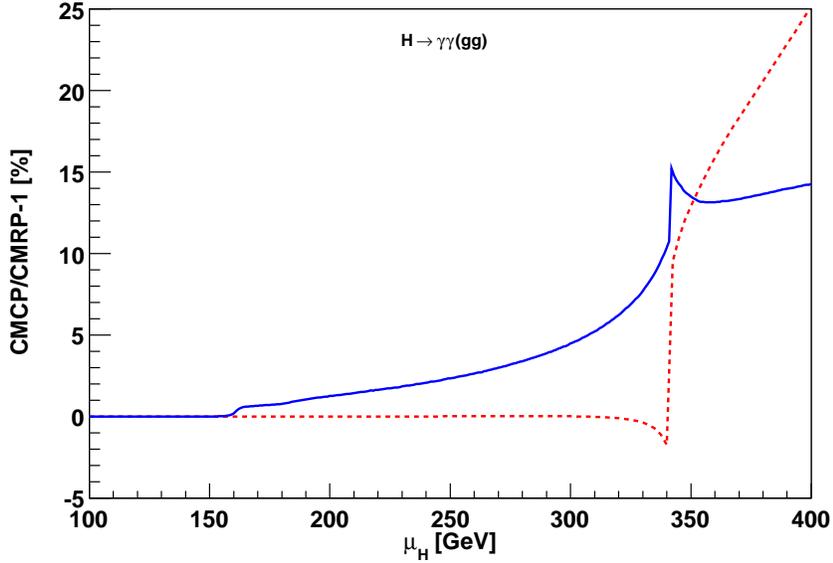}
\end{center}
\vspace{-0.5cm}
\caption[]{\label{c_dev}Comparison of predictions in the CMRP 
and CMCP schemes for the decays $H \to \gamma \gamma$ (blue, solid line) and
for $H \to gg$ (red, dashed line). See \sect{schemes} for the scheme definitions.}
\end{figure}

\begin{figure}[ht!]
\begin{center} 
\includegraphics[bb=0 0 567 384,width=12cm]{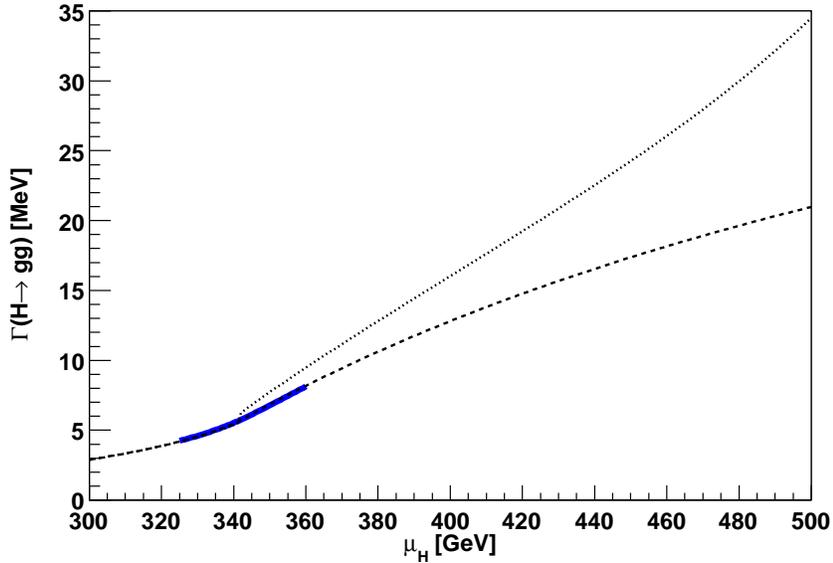}
\end{center}
\vspace{-0.5cm}
\caption[]{\label{G_hgg}Comparison of the decay width $\Gamma(H \to gg)$
  in the CMRP (dashed line) and the CMCP (dotted line) scheme in the high mass
  region.  The effect of a complex top quark pole in CMCP (with a top total,
  on-shell, width of $13.1\,$GeV) is given by the blue, solid line. See
  \sect{schemes} for the scheme definitions.}
\end{figure}

\begin{figure}[ht!]
\begin{center} 
\includegraphics[bb=0 0 567 384,width=12cm]{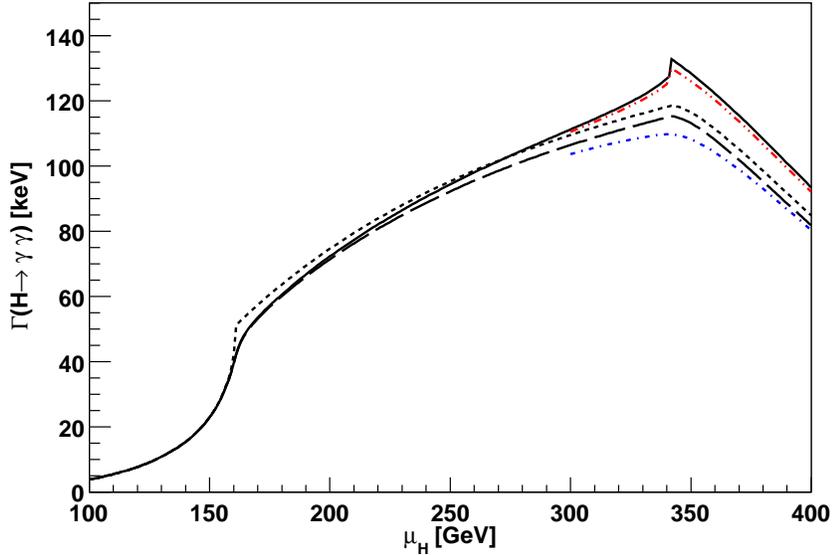}
\end{center}
\vspace{-0.5cm}
\caption[]{\label{G_hgamgam}Comparison of the decay width $\Gamma(H \to
  \gamma\gamma)$ in the RMRP (dotted line), the CMRP (dashed line) and the
  CMCP (solid line) scheme.  The effect of a complex top quark pole in CMCP
  (with a top total, on-shell, width of $13.1\,$GeV) is given by the
  blue, dash-dotted line. The red, dash-double-dotted line corresponds
  to a top width of $1.31\,$GeV. See \sect{schemes} for the scheme
  definitions.}
\end{figure}

\noindent
A detailed comparison of predictions in the CMRP and CMCP schemes for 
$H \to \gamma \gamma , gg$ is shown in \fig{c_dev}. The partial decay width 
$H \to gg$ is shown in \fig{G_hgg} where we compare the CMRP(=RMRP) and CMCP 
schemes.
Similarly we compare the partial decay width $H \to \gamma\gamma$ 
in the RMRP, CMRP and CMCP schemes in \fig{G_hgamgam}.

The relatively large effects in $\Gamma(H \to \gamma\gamma)$ or 
$\Gamma(H \to gg)$ at large values of $\srph$ are still compatible with
the naive $\ord{\lgh/\srph}$ argument. Consider \fig{c_dev}
for $\Gamma(H \to \gamma\gamma)$; for instance, at $\srph = 365\,$GeV
we have $\lgh/\srph= 0.065$ with a variation between the CMRP and
CMCP schemes of $13.3\%$ giving a correction factor of $2\,\lgh/\srph$.
Clearly, the large increase in the Higgs boson width for increasing values of
$\srph$ makes it questionable to use a perturbative description for
the Higgs-resonant part of $pp \to X$ when we have a very heavy Higgs boson.

\begin{figure}[t!]
\begin{center} 
\includegraphics[bb=0 0 567 384,width=12cm]{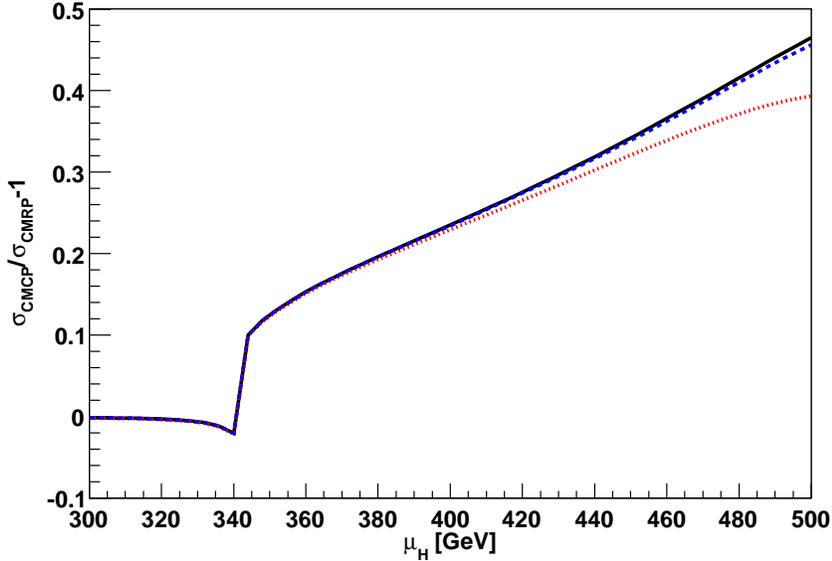}
\end{center}
\vspace{-0.5cm}
\caption[]{\label{R_ppH}The ratio $\sigma_{\CMCP}/\sigma_{\CMRP}$ for
  the production cross section $pp \to H$, as a function of $\srph$, for
  different energies, $\sqrt{s} = 3\,$TeV (red, dotted line), $\sqrt{s} =
  10\,$TeV (blue, dashed line) and $\sqrt{s} = 14\,$TeV (black, solid
  line). The cross sections are computed with MSTW2008 LO PDFs with 
  factorization scale $\mu_{\ssF}=\srph$ for CMRP and $\mu_{\ssF}=\sbrph$ 
  for CMCP.}
\end{figure}

The relevance of this result is clear: if a light Higgs boson is not discovered, 
one of the goals of LHC will be to exclude a standard model Higgs up 
to $600\,$GeV~\cite{EP600}; already at $500\,$GeV we have
\bq
\frac{\sigma_{\CMCP}(gg \to H)}{\sigma_{\CMRP}(gg \to H)} = 1.64,
\qquad \hbox{parton level},
\eq
comparable to the effect of NLO QCD corrections. We have computed also
$\sigma(pp \to H)$ in the two schemes using MSTW 2008 LO parton
distribution functions (PDF)~\cite{Martin:2009iq}.
The ratio is given in \fig{R_ppH}, for different values of $s$; in this
figure we present
\bq
\frac{\sigma_{\CMCP}(pp \to H)}{\sigma_{\CMRP}(pp \to H)},
\eq
where the numerator is evaluated at $\sbrph$ (\eqn{MHdef}) while 
the denominator corresponds to $\srph$. Here we use
\bq
\sigma(pp \to H) = \sigma_0\,\tau_{\ssH}\,\frac{dL^{gg}}{d \tau_{\ssH}},
\qquad
\frac{dL^{gg}}{d \tau} = \int_{\tau}^1\,\frac{dx}{x}\,
g\lpar x,\mu_{\ssF}\rpar\,g\lpar \frac{\tau}{x},\mu_{\ssF}\rpar,
\eq
where $\tau = \rph(\brph)/s$, $\sigma_0$ is the parton 
level cross section and $g$ is the gluon PDF (with factorization scale
$\mu_{\ssF}=\srph$ for CMRP and $\mu_{\ssF}=\sbrph$ for CMCP).

Note that this ratio is only an indicator of the (large) size of the effect 
since, for a realistic value of the cross section, NLO(NNLO) QCD corrections 
should be included, see Ref.~\cite{deFlorian:2009hc} for updated cross sections 
at the Tevatron and the LHC.

\begin{figure}[ht]
\begin{center} 
\includegraphics[bb=0 0 567 384,width=12cm]{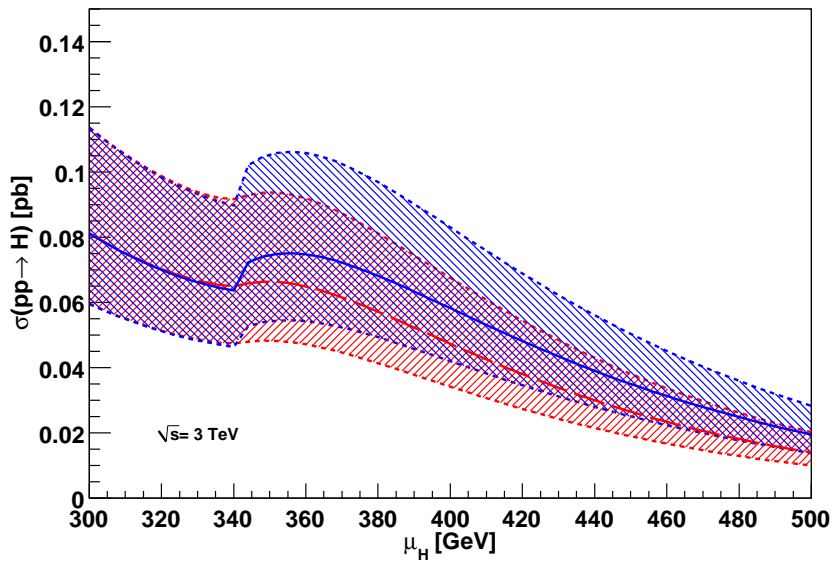}
\end{center}
\vspace{-0.5cm}
\caption[]{\label{S_ppH_3T}The production cross section $pp \to H$ at
$\sqrt{s}= 3\,$TeV for CMRP (red, wide-dashed line) and CMCP (blue, solid
line). The shaded areas surrounded by the dashed lines give the scale
uncertainty obtained by varying $\srph/2 < \mu_{\ssR} = \mu_{\ssF} <
2\,\srph$ in the CMRP scheme and $\sbrph/2 < \mu_{\ssR} = \mu_{\ssF} <
2\,\sbrph$ (\eqn{MHdef}) in the CMCP scheme. We have used MSTW2008 LO
PDFs.}
\end{figure}

In \fig{S_ppH_3T} we show the corresponding cross section for
$\sqrt{s}= 3\,$TeV, including an estimate of the uncertainty induced
by varying renormalization and factorization scales (kept equal for simplicity);
$\srph/2 \le \mu_{\ssR} = \mu_{\ssF} \le 2\,\srph$ for
CMRP and $\sbrph/2 \le \mu_{\ssR} = \mu_{\ssF} \le 2\,\sbrph$ for
CMCP ($\sbrph$ is given in \eqn{MHdef}). In \fig{S_ppH_3T} we do not include 
the uncertainty associated to PDFs. 

\begin{figure}[ht]
\begin{center} 
\includegraphics[bb=0 0 567 384,width=12cm]{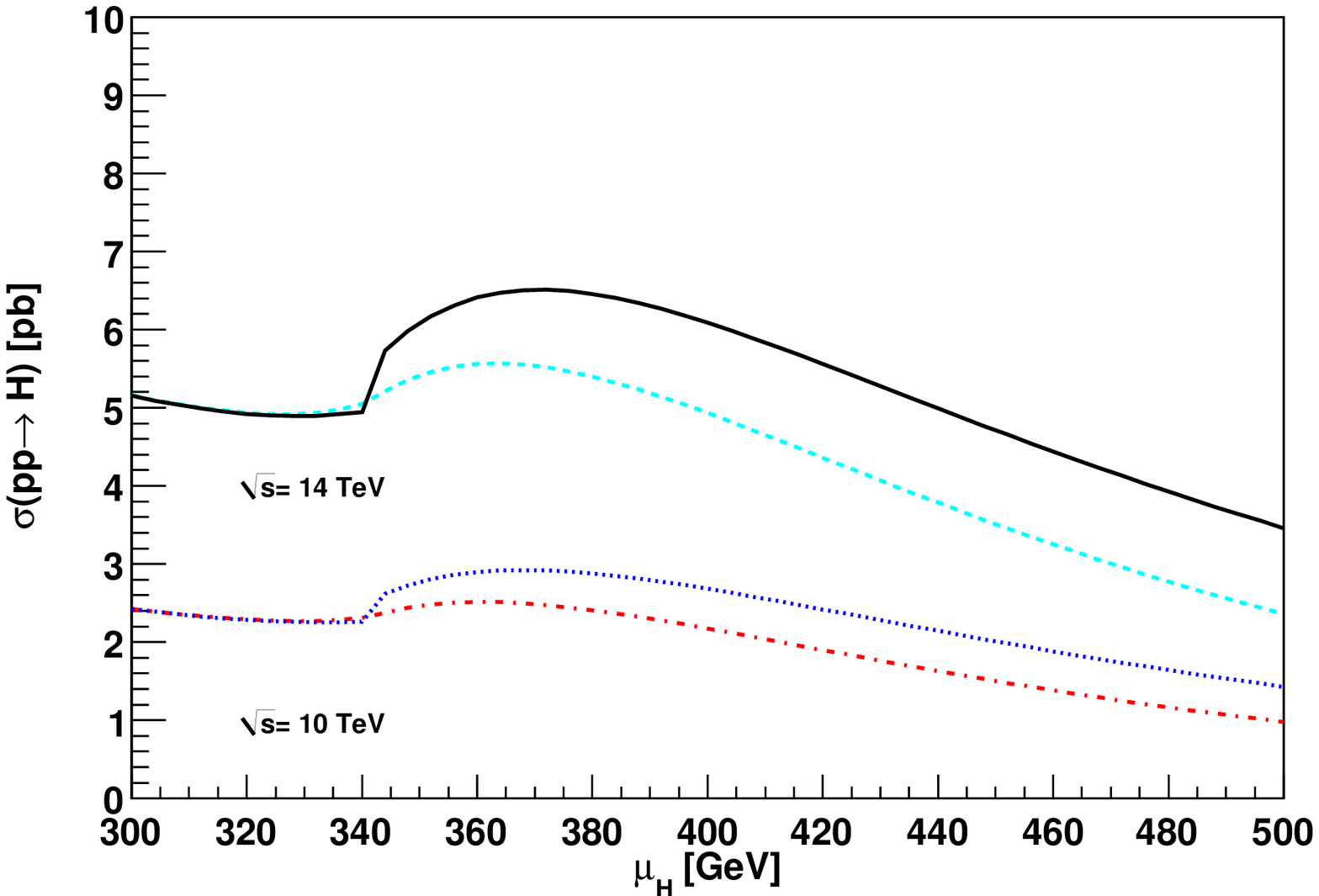}
\end{center}
\vspace{-0.5cm}
\caption[]{The production cross section $pp \to H$ at $\sqrt{s}=
  10\,$TeV for CMRP (red, dashed-dotted line) and CMCP (blue, dotted
  line); at $\sqrt{s} = 14\,$TeV for CMRP (cyan, dashed line) and CMCP
  (black, solid line). We have used MSTW2008 LO PDFs with factorization 
  scale $\mu_{\ssF}=\srph$ for CMRP and $\mu_{\ssF}=\sbrph$ for CMCP.}
\label{S_ppH_10a14T}
\end{figure}

The ratio between the two cross sections is stable under these 
variations, e.g. is between $1.4546$ and $1.4574$ at $\sqrt{s}= 10\,$TeV
and $\srph = 500\,$GeV. The production cross sections for $\sqrt{s} = 10, 14\,$
TeV are shown in \fig{S_ppH_10a14T}; note that here we use
$\alpha_{\ssS}(M_{\ssZ}) = 0.13934$ to be consistent with the LO
PDFs~\cite{Martin:2009iq}. 

\begin{figure}[h!]
\begin{center} 
\includegraphics[bb=0 0 567 547,width=9cm]{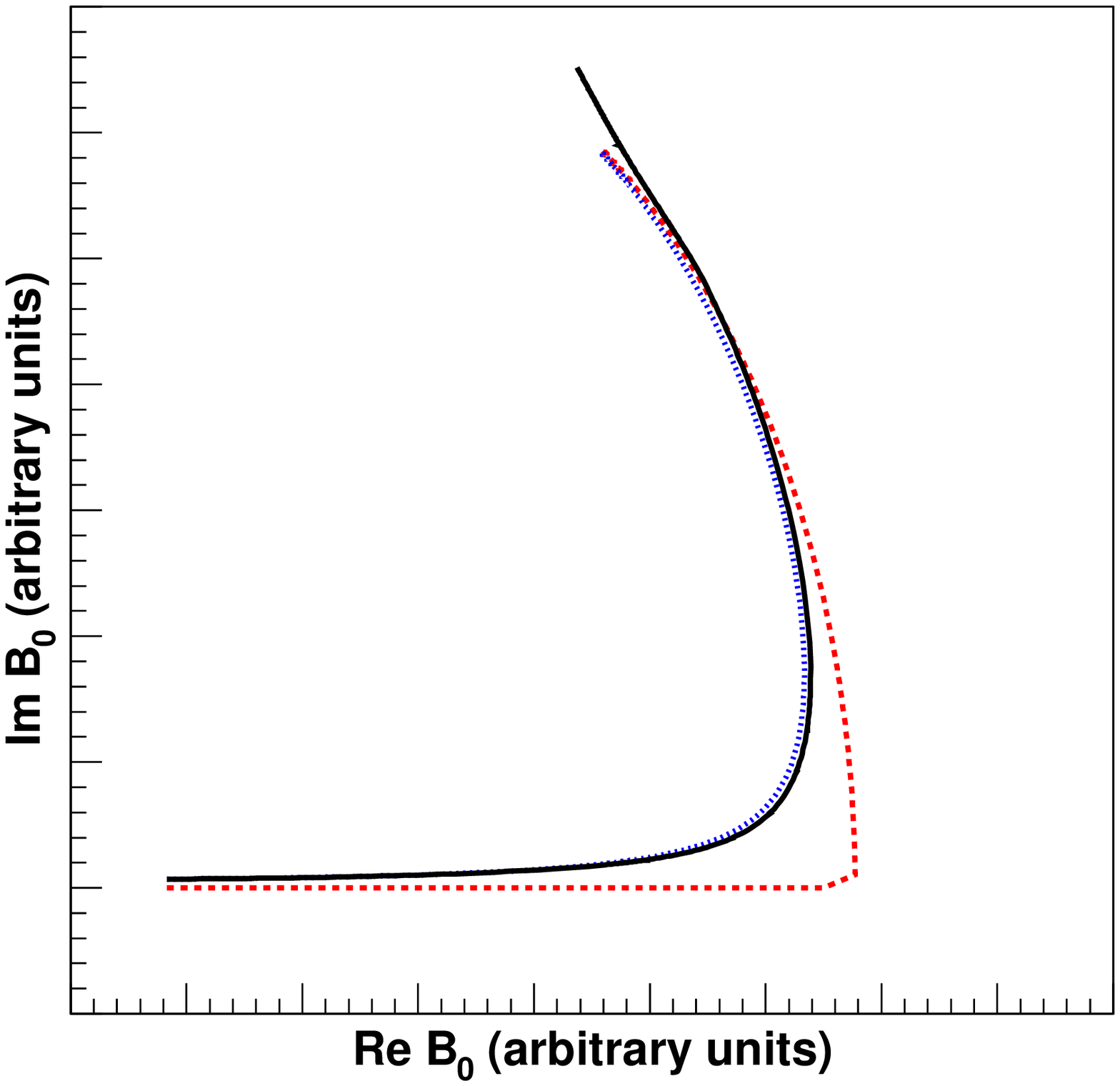}
\end{center}
\vspace{-0.5cm}
\caption[]{\label{B0}The scalar two-point function with two real
internal masses (RMRP scheme) $M^2_{\ssZ}$ or two complex internal
masses (CMRP - CMCP schemes) $s_{\ssZ}$.  $B_0$ is computed in the RMRP
(red dash-line), CMRP (blue dot-line) and CMCP (black solid-line) schemes. 
The Mandelstam invariant is
$\rph$ in the RMRP - CMRP schemes and $\cph$ in the CMCP scheme, with
$100\,\GeV < \srph < 500\,\GeV$. See \sect{schemes} for the scheme
definitions.}
\end{figure}

In order to better understand the numerical differences in the three schemes we show in 
\fig{B0} a scalar two-point function with two internal $Z$ masses. 
In the RMRP scheme both the internal masses and the Mandelstam invariant
$\rph$ are real; in the CMRP scheme we replace $M^2_{\ssZ}$ with the
corresponding complex pole, $s_{\ssZ}$; finally, in the CMCP scheme,
also the invariant becomes complex and equal to $\cph$. In \fig{B0} we
vary $\srph$ between $100\,$GeV and $500\,$GeV where $\lgh =
146.89\,$GeV is huge; here deviations between CMRP and CMCP become
sizable. This simple example shows the general features of one-loop
corrections in the three schemes; CMRP - CMCP smoothly interpolate the
RMRP results around normal thresholds and, when $\rph$ becomes larger
and larger, CMCP starts deviating from RMRP - CMRP.

\begin{figure}[h!]
\begin{center}
\includegraphics[bb=0 0 567 384,width=12cm]{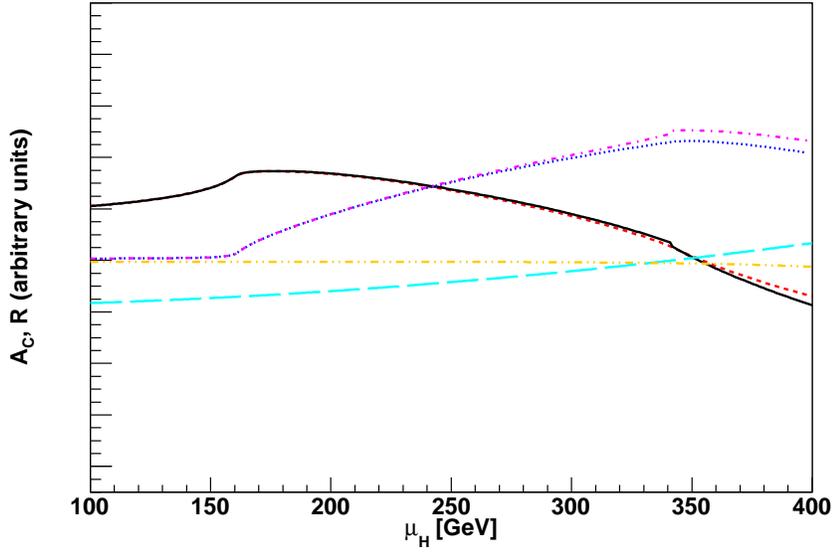}
\end{center}
\vspace{-0.5cm}
\caption[]{\label{C0rat}The $C_0$ part of the amplitude for $H \to
  \gamma \gamma$ and the corresponding rational term of \eqn{CpRat} with
  $100\,\GeV < \srph < 300\,\GeV$. The black, solid (CMCP) and red, dashed
  (CMRP) lines give the real part whereas blue, dotted (CMRP) and magenta,
  dash-dotted (CMCP) lines give the imaginary part. The cyan, wide-dashed
  (orange, dash-double-dotted) line gives the rational real(imaginary)
  part of the amplitude without appreciable differences between the
  schemes. The imaginary part of the rational term is always small and
  negligible. See \sect{schemes} for the scheme definitions.}

\end{figure}

We also consider the expression for the amplitude $A\lpar H \to 
\gamma \gamma\rpar$ which can be split into a part containing only 
$C_0\,$-functions and a rational term. We write
\bq
\Gamma\lpar H \to \gamma\gamma\rpar =
\frac{\alpha^2\,\gf}{32\,\sqrt{2}\,\pi^3}\,
\frac{\mid s_{\ssW}\mid^2}{\sbrph}\,\bmid A\bmid^2.
\eq
If we introduce auxiliary variables,
\bq
\brph= \srph\,\lpar \rph + \gamma^2_{\ssH}\rpar^{1/2}, \quad
x_t= \frac{m^2_t}{\brph}, \quad
x_{\ssH}= \frac{s_{\ssH}}{\brph}, \quad
x_{\ssW}= \frac{s_{\ssW}}{\brph},
\eq
the amplitude can be written as $A= A_{\ssC} + R$ where
\bqa
A_{\ssC} &=& -\,\frac{8}{3}\,x_t\,\frac{x_{\ssH} - 4\,x_t}{x_{\ssW}}\,
C_0\lpar 0\,,\,0\,,\,-s_{\ssH}\,,\,m_t\,,\,m_t\,,\,m_t \rpar 
+ 6\,\lpar x_{\ssH} - 2\,x_{\ssW} \rpar\,
C_0\lpar 0\,,\,0\,,\,-s_{\ssH}\,,\,s_{\ssW}\,,\,s_{\ssW}\,,\,s_{\ssW} \rpar,
\nl
R &=& -\,\frac{16}{3}\,\frac{x_t}{x_{\ssW}} + \frac{x_{\ssH}}{x_{\ssW}} + 6. 
\label{CpRat}
\eqa
A comparison for the real and imaginary parts in the CMRP and CMCP schemes is 
shown in \fig{C0rat}.

\begin{itemize}
\item $\mathbf{H \to \barb b}$ 
\end{itemize} 

\begin{figure}[h!]
\begin{center} 
\includegraphics[bb=0 0 567 384,width=12cm]{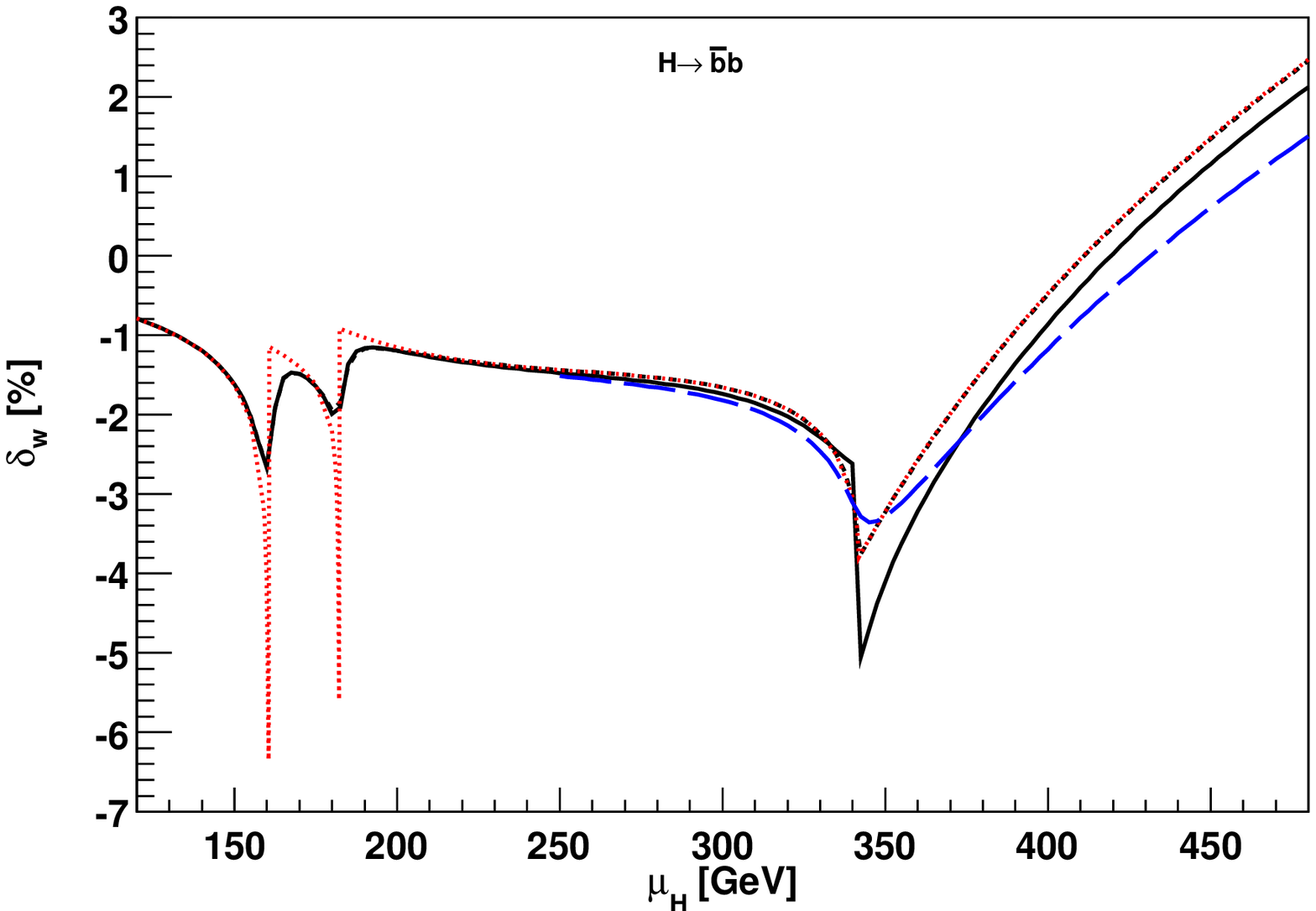}
\end{center}
\vspace{-0.5cm}
\caption[]{\label{hbb_w} The weak one-loop radiative corrections to $H \to
\barb b$ in the RMRP scheme (red, dotted line), in the CMRP scheme (black,
dashed line) and in the CMCP scheme (black, solid line). The effect of a
complex top quark pole in CMCP (with a top total, on-shell, width of
$13.1\,$GeV) is given by the blue wide-dashed line.  See \sect{schemes} for
the scheme definitions.  The result corresponding to $\Gamma_t
=\Gamma^{\NLO}_t = 1.31\,$GeV has no appreciable difference compared to the one
at $\Gamma_t = 0$.}
\end{figure}

\begin{figure}[h!]
\begin{center} 
\includegraphics[bb=0 0 567 384,width=12cm]{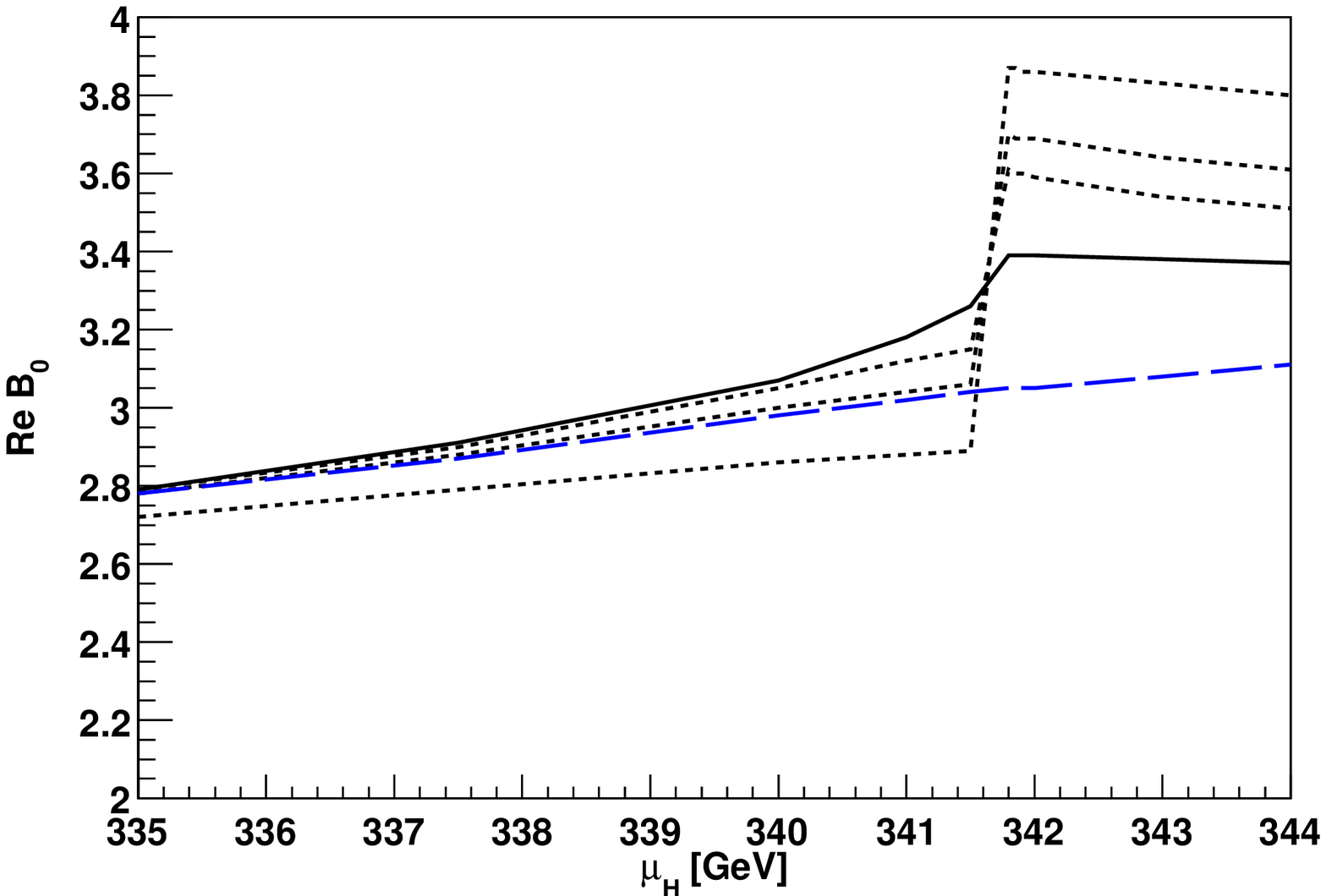}
\end{center}
\vspace{-0.5cm}
\caption[]{\label{tt_threshold}The $\Reb\,B_0(-\cph\,;\,m_t,m_t)$ around the
$\bart t$ threshold. The solid line corresponds to a real Higgs mass, $\lgh =
0$, whereas dashed lines correspond to increasing values of $\lgh$. The
blue, wide-dashed line corresponds to a finite $\lgh$ and to a complex top quark pole
(with an on-shell width of $13.1\,$GeV).}
\end{figure}

\noindent
Results for $H \to \barb b$ are shown in \tabn{THbbWwide} for the pure  
weak percentage one-loop corrections in the three schemes and for a wide 
range of Higgs masses;
\begin{table}[ht]\centering
\setlength{\arraycolsep}{\tabcolsep}
\renewcommand\arraystretch{0.9}
\begin{tabular}{|c|c|c|c|c|}
\hline 
&&&& \\
$\mh\;$[GeV] & 
$\delta_{\ssW}$ RMRP [$\%$]  & $\delta_{\ssW}$ CMRP [$\%$] & 
$\delta_{\ssW}$ CMCP [$\%$] & $\delta_{\ssW}$ CMCP [$\%$] \\
 & massless & massless & massless & massive \\
&&&& \\
\hline
&&&& \\
$   120 $ & $   -0.7890 $ & $  -0.7904 $ & $   -0.7904$ & $ -0.7837$ \\
$   130 $ & $   -0.9557 $ & $  -0.9572 $ & $   -0.9573$ & $ -0.9509$ \\
$   140 $ & $   -1.1978 $ & $  -1.1986 $ & $   -1.1986$ & $ -1.1922$ \\
$   150 $ & $   -1.6215 $ & $  -1.6146 $ & $   -1.6149$ & $ -1.6078$ \\
$   160 $ & $   -4.2656 $ & $  -2.6458 $ & $   -2.6690$ & $ -2.6587$ \\
$   170 $ & $   -1.3987 $ & $  -1.4914 $ & $   -1.4875$ & $ -1.4822$ \\
$   180 $ & $   -2.1989 $ & $  -1.9435 $ & $   -1.9912$ & $ -1.9858$ \\
$   190 $ & $   -1.0338 $ & $  -1.1744 $ & $   -1.1590$ & $ -1.1569$ \\
$   200 $ & $   -1.1547 $ & $  -1.1967 $ & $   -1.1987$ & $ -1.1974$ \\
$   210 $ & $   -1.2452 $ & $  -1.2621 $ & $   -1.2730$ & $ -1.2723$ \\
$   220 $ & $   -1.3132 $ & $  -1.3198 $ & $   -1.3379$ & $ -1.3376$ \\
$   230 $ & $   -1.3647 $ & $  -1.3665 $ & $   -1.3914$ & $ -1.3917$ \\
$   240 $ & $   -1.4047 $ & $  -1.4044 $ & $   -1.4363$ & $ -1.4370$ \\
$   250 $ & $   -1.4376 $ & $  -1.4365 $ & $   -1.4759$ & $ -1.4769$ \\
$   260 $ & $   -1.4674 $ & $  -1.4665 $ & $   -1.5138$ & $ -1.5151$ \\
$   270 $ & $   -1.4985 $ & $  -1.4981 $ & $   -1.5539$ & $ -1.5555$ \\
$   280 $ & $   -1.5357 $ & $  -1.5361 $ & $   -1.6008$ & $ -1.6026$ \\
$   290 $ & $   -1.5851 $ & $  -1.5865 $ & $   -1.6604$ & $ -1.6624$ \\
$   300 $ & $   -1.6557 $ & $  -1.6582 $ & $   -1.7410$ & $ -1.7431$ \\
$   400 $ & $   -0.4736 $ & $  -0.4865 $ & $   -0.8589$ & $ -0.8633$ \\
$   450 $ & $   +1.4855 $ & $  +1.4687 $ & $   +1.1579$ & $ +1.1517$ \\
&&&& \\
\hline
\end{tabular}
\caption[] {Percentage one-loop pure weak corrections for $H \to \barb
b$; the first entry is the RMRP scheme, the second entry is the CMRP
scheme, the third entry is the CMCP scheme while the fourth entry is the
CMCP scheme with finite $m_b$ (\sect{schemes}).}
\label{THbbWwide}
\end{table}
%
in \tabn{THbbwide} the electroweak (weak + QED) percentage one-loop 
corrections are given.
\begin{table}[ht]\centering
\setlength{\arraycolsep}{\tabcolsep}
\renewcommand\arraystretch{0.9}
\begin{tabular}{|c|c|c|}
\hline 
&& \\
$\mh\;$[GeV] & 
$\delta_{\EW}$ CMRP [$\%$] & 
$\delta_{\EW}$ CMCP [$\%$] \\
&& \\
\hline
&& \\
$   120 $ & $   -0.9728 $ & $  -0.9729$ \\
$   130 $ & $   -1.1467 $ & $  -1.1468$ \\
$   140 $ & $   -1.3941 $ & $  -1.3942$ \\
$   150 $ & $   -1.8151 $ & $  -1.8154$ \\
$   160 $ & $   -2.8485 $ & $  -2.8716$ \\
$   170 $ & $   -1.7039 $ & $  -1.7000$ \\
$   180 $ & $   -2.1606 $ & $  -2.2083$ \\
$   190 $ & $   -1.3990 $ & $  -1.3837$ \\
$   200 $ & $   -1.4262 $ & $  -1.4283$ \\
$   210 $ & $   -1.4961 $ & $  -1.5071$ \\
$   220 $ & $   -1.5580 $ & $  -1.5761$ \\
$   230 $ & $   -1.6087 $ & $  -1.6337$ \\
$   240 $ & $   -1.6503 $ & $  -1.6824$ \\
$   250 $ & $   -1.6861 $ & $  -1.7256$ \\
$   260 $ & $   -1.7194 $ & $  -1.7669$ \\
$   270 $ & $   -1.7543 $ & $  -1.8102$ \\
$   280 $ & $   -1.7953 $ & $  -1.8602$ \\
$   290 $ & $   -1.8487 $ & $  -1.9228$ \\
$   300 $ & $   -1.9232 $ & $  -2.0061$ \\
$   400 $ & $   -0.7754 $ & $  -1.1488$ \\
$   450 $ & $   +1.1697 $ & $  +0.8571$ \\
&& \\
\hline
\end{tabular}
\caption[] {Percentage one-loop electroweak (weak + QED) corrections for 
$H \to \barb b$ with a massive b-quark; the first entry is the CMRP 
scheme, the second entry is the CMCP scheme (\sect{schemes}).}
\label{THbbwide}
\end{table}
%
For weak corrections the results for $H \to \barb b$ are presented
graphically in \fig{hbb_w}; 
the figure shows cusps at the $\bart t\,$-threshold due to the fact that
the top quark mass is kept real. The origin of the cusps is in a $B_0\,$-function
with $p^2$ fixed at the complex Higgs pole. The size of the cusps can be related
to the large Higgs width at the $\bart t\,$-threshold as illustrated in
\fig{tt_threshold} where we analyze $\Reb\,B_0(-\cph\,;\,m_t,m_t)$ around 
the $\bart t$ threshold. Here the solid line corresponds to $\lgh = 0$, 
whereas dash-lines correspond to increasing values of $\lgh$. 
As one can see the limit $\lgh \to 0$ is continuous and there is no
artifact due to the analytical continuation.
The wide-dashed blue line of \fig{tt_threshold} corresponds to a finite value of
$\lgh$ and to a complex top pole (with an on-shell width of
$13.1\,$GeV). 

As it is evident the introduction of a complex top quark pole
completely cures the shape of the corrections. Therefore, this gives further 
evidence to using the CMCP scheme, at least from a theoretical point of view (the 
top quark total width is, unfortunately, poorly known). It is worth noting that the
numerical effect given by the blue curve should be interpreted as an upper bound 
on the effect of a top quark complex pole since the experimental result is an 
upper bound, $\Gamma_t < 13.1\,$GeV at $95\%$ C.L. Note that, from theory it 
follows $\Gamma^{\LO}_t = 1.47\,$GeV and $\Gamma^{\NLO}_t= 1.31\,$GeV. 

Finally, we observe that the $\ord{\lgh/\srph}$ effects, which can
reach several percent at large values of $\srph$, have a modest effect on 
all those processes which start at $\ord{g^2}$ (the effect being on NLO 
corrections) whereas the effect is considerably larger for those processes, e.g. 
$H \to \gamma \gamma (gg)$, which start directly at NLO ($\ord{g^6}$).
\section{Conclusions \label{Conclu}}
In this paper we have shown how to continue Feynman integrals into the second 
Riemann sheet, in a way that can be easily implemented in any program aimed to
compute pseudo-observables related to Higgs physics at Tevatron and LHC.
Pseudo-observables give, in a natural way, the possibility of translating
experimental data into a language that has a direct connection to 
unambiguous theoretical calculations. Using our framework one can freely
compute quantities (otherwise non-existing) like Higgs production cross section
and Higgs partial decay widths. 

An unstable particle cannot belong to the in/out basis of the Hilbert space, 
nevertheless concepts like production or decay of an unstable particle becomes 
aliases for pseudo-observables that have a well defined meaning and a direct 
relation to measured data. 

In this paper a new scheme is introduced which is the (complete) complex mass 
scheme with complex, external Higgs boson (or, equivalently, any other external
unstable particle) where the LSZ procedure is carried out at the Higgs complex pole 
(on the second Riemann sheet).

Pseudo-observables have been a very useful concept at LEP 
(e.g. Ref.\cite{Grunewald:2000ju})  and will continue to play an important role 
at LHC, although more difficult if deviations from the SM will emerge; in this 
case model independent approaches are required allowing for the extraction of useful 
quantities that can be fitted with different models.
 
The usual objection against moving standard model Higgs pseudo-observables into 
the second Riemann sheet of the $S\,$-matrix is that a light Higgs boson, say 
below $140\,$GeV, has a very narrow width and the effects induced are tiny. 
Admittedly, it is a well taken point for all practical consequences but
one should remember that the Higgs boson width rapidly increases after 
the opening of the $WW$ and $ZZ$ channels and, because of this, the on-shell
treatment of an external Higgs particle becomes inadequate as a description of data
if the Higgs is not (very) light. 

Furthermore, most of the experimental plots concerning Higgs physics extend well 
above, say, $200\,$GeV and, if a light Higgs boson is not discovered, one of the 
goals of LHC will be to exclude a standard model Higgs up to $600\,$GeV (see 
Ref.~\cite{EP600} for an exclusion plot of the SM Higgs boson for the various 
channels as well as the combination for masses up to $600\,$GeV). Already at 
$500\,$GeV the ratio CMCP/CMRP for the $gg \to H$ cross section is large and 
comparable to higher order QCD corrections.

On top of all practical implications one should admit that it is hard to sustain 
a wrong theoretical description of experimental data if the correct one is 
available, independently on the size of the effect.

Finally, our results show that, above the $\bart t\,$-threshold the Higgs-resonant
contribution to $pp \to X$, correctly described in the CMCP scheme, is strongly
influenced by the large imaginary part of the Higgs complex pole and the use
of the conventional on-shell description of Higgs pseudo-observables becomes highly 
questionable, even from a numerical point of view.
\Acknowledgments
G.P. is indebted to Stefan Dittmaier for an important discussion
on $H \to 4\,$f. 
We gratefully acknowledge several discussions with W.~Hollik and R.~Pittau.
\clearpage
\begin{appendix}

\end{appendix}
\end{document}